\PassOptionsToPackage{unicode}{hyperref}
\PassOptionsToPackage{hyphens}{url}
\documentclass[
  11pt,
]{article}
\usepackage{amsmath,amssymb}
\usepackage{setspace}
\usepackage{iftex}
\ifPDFTeX
  \usepackage[T1]{fontenc}
  \usepackage[utf8]{inputenc}
  \usepackage{textcomp} 
\else 
  \usepackage{unicode-math} 
  \defaultfontfeatures{Scale=MatchLowercase}
  \defaultfontfeatures[\rmfamily]{Ligatures=TeX,Scale=1}
\fi
\usepackage[]{mathptmx}
\ifPDFTeX\else
\fi
\IfFileExists{upquote.sty}{\usepackage{upquote}}{}
\IfFileExists{microtype.sty}{
  \usepackage[]{microtype}
  \UseMicrotypeSet[protrusion]{basicmath} 
}{}
\makeatletter
\@ifundefined{KOMAClassName}{
  \IfFileExists{parskip.sty}{%
    \usepackage{parskip}
  }{
    \setlength{\parindent}{0pt}
    \setlength{\parskip}{6pt plus 2pt minus 1pt}}
}{
  \KOMAoptions{parskip=half}}
\makeatother
\usepackage{xcolor}
\usepackage[margin=2.4cm]{geometry}
\usepackage{longtable,booktabs,array}
\usepackage{calc} 
\usepackage{etoolbox}
\makeatletter
\patchcmd\longtable{\par}{\if@noskipsec\mbox{}\fi\par}{}{}
\makeatother
\IfFileExists{footnotehyper.sty}{\usepackage{footnotehyper}}{\usepackage{footnote}}
\makesavenoteenv{longtable}
\providecommand{\tightlist}{%
  \setlength{\itemsep}{0pt}\setlength{\parskip}{0pt}}
\usepackage{amsmath,amssymb}
\usepackage{courier}
\usepackage{graphicx,float}
\usepackage{longtable,booktabs,array}
\AtBeginEnvironment{longtable}{\footnotesize}
\usepackage{titlesec}
\titleformat{\section}{\normalfont\large\bfseries}{\thesection}{0em}{}
\titleformat{\subsection}{\normalfont\normalsize\bfseries}{\thesubsection}{0em}{}
\ifLuaTeX
  \usepackage{selnolig}  
\fi
\IfFileExists{bookmark.sty}{\usepackage{bookmark}}{\usepackage{hyperref}}
\IfFileExists{xurl.sty}{\usepackage{xurl}}{} 
\hypersetup{
  pdftitle={Prime factorisation of stable-matching instances: uniqueness, simultaneous products, and an exact census},
  pdfauthor={Yoshiteru Ishida},
  hidelinks,
  pdfcreator={LaTeX via pandoc}}

\title{Prime factorisation of stable-matching instances: uniqueness,
simultaneous products, and an exact census}
\author{Yoshiteru Ishida\thanks{Toyohashi University of Technology, Toyohashi, Japan. \texttt{ishida@cs.tut.ac.jp}. ORCID 0000-0003-1641-5385.}}
\date{28 September 2026}

\begin{document}
\maketitle

\setstretch{1.05}
\hypertarget{abstract}{%
\section{Abstract}\label{abstract}}

Every balanced instance of the stable marriage problem with strict
complete preferences has a unique finest partition into prime blocks,
and that single partition simultaneously factors three different
structures: the reachable execution digraph as a Cartesian product, the
proposal-prefix antimatroid as a direct sum, and the stable-matching
lattice as a direct product. The converse fails, and fails at every size
from two on: two explicit families share the identical Boolean-cube
execution while one is maximally decomposable with a single stable
matching and the other is prime with n.~Uniqueness yields an exact
census, a recursion counting the prime instances at every size, under
which exactly 88,478,208 of the 110,075,314,176 profiles with four
agents on each side are decomposable and the decomposable fraction is
asymptotically n! / n\textsuperscript{2n}. The blocks are characterised
as the square components of the mutual-rank filtration, so the partition
is computable in polynomial time and the factorisation is a tool rather
than only a fact.

\hypertarget{introduction}{%
\section{1. Introduction}\label{introduction}}

Deferred acceptance is normally analysed through its terminal stable
matching. This paper analyses the instance instead, and asks the
question one asks of an integer: can it be broken into smaller pieces,
is the breaking unique, and what is preserved by it.

The answer is that it can, that it is, and that a surprising amount is
preserved. An instance decomposes when its agents split into paired
blocks of proposers and receivers, each block mutually top-ranked within
itself. Such a partition always has a unique finest refinement (Theorem
6), and the blocks of that refinement are the prime instances of the
title. What makes the decomposition worth having is not the uniqueness
alone but what one partition does at once (Theorem 7): the execution
digraph becomes a Cartesian product over the blocks, the antimatroid of
feasible proposal prefixes becomes a direct sum, and the lattice of
stable matchings becomes a direct product. Three structures that are
otherwise studied separately are factored by the same object.

That makes the theorem a reduction tool. Any quantity that composes over
a product - the state count, the proposal total, the number of stable
matchings, the commutation index - is determined by its values on prime
blocks, and Corollary 7.2 sorts those quantities into the multiplicative
ones and the additive ones. To compute such a quantity it suffices to
find the finest partition, which Theorem 9.4 reduces to reading off the
square components of a filtration of bipartite graphs, and this can be
done in polynomial time.

The limits of the tool are as sharp as the tool. Behavioural
independence - the execution splitting as a direct sum - does not imply
input independence (Theorem 8), and the failure is not exceptional:
Proposition 12.3 exhibits two families, one at every size, that share
the identical Boolean-cube execution while sitting at opposite ends of
both of the things the execution cannot see. Section 5 makes that
precise, and Proposition 9.10 shows that the behavioural decomposition,
unlike the input one, carries no nesting at all: its factorisations form
an interval in the partition lattice and their number is a Bell number
depending on nothing but how many prime factors there are.

Finally, uniqueness is what makes counting possible. Theorem 9 turns the
finest partition into a recursion for the number of prime instances at
every size, with exact values at small sizes and an asymptotic
decomposable fraction of n! / n\textsuperscript{2n}. The small values
are given in full so that a reader can check an independent
implementation against them; that is the part of a structure theory that
makes it usable by other people.

\emph{Related work.} Deferred acceptance and the lattice of stable
matchings are classical {[}1,2,3{]}. The structure of what an algorithm
is able to do next is the greedoid tradition of Korte, Lovasz and
Schrader {[}4{]}; antimatroids induced by matchings are studied by
Kawase and Yamaguchi {[}5{]}, and which preference profiles determine
the proposals the Gale-Shapley algorithm makes is the question of
Sukegawa and Yamamoto {[}6{]}. A uniqueness statement of the kind proved
here was first isolated for lattices by Dilworth {[}7{]} and for graphs
under the Cartesian product by Sabidussi {[}8{]}, with later extensions
to graphs with loops {[}9{]}; convex geometries, the duals of
antimatroids, are due to Edelman and Jamison {[}10{]}. That independent
proposals commute is the phenomenon trace theory {[}11{]} and Keller's
parallel program schemata {[}12{]} axiomatise, and Section 4 says
exactly how far commutation alone goes. Bounds on the number of stable
matchings are due to Karlin, Oveis Gharan and Weber {[}13{]} from above,
to Benjamin, Converse and Krieger {[}14{]} from below, and at small
sizes to Thurber {[}15{]}. Order-independence of the kind Section 2
proves is the defining property of abelian networks {[}16{]} and was
established for threshold dynamics by Goles and Olivos {[}17{]}. Closest
to the subject of this paper, the decomposability of a single instance
was studied by Veklerov {[}18{]}, whose first criterion is the static
condition used here, and the Latin instances - those in which the two
ranks of a pair sum to n + 1 - go back to Knuth {[}19{]} and are the
latin marriages of {[}14{]}. The paper continues the author's {[}20{]},
which gives the operator formulation of the run used in Section 2, and
{[}21{]}, whose regular anti-phase templates supply the two families of
Section 5.

\emph{Relation to the author's monograph.} Some of the material below
was presented without proof, through worked examples rather than
statements, in the author's monograph {[}22{]}. Its Chapter 3 classifies
the size-three instances by diagrams and introduces the decomposition of
an instance informally, Chapter 4 studies the Latin instances of
{[}19,14{]}, Chapter 5 introduces a cyclic family with two stable
matchings, and Chapter 7 states the block condition used here in the
form of {[}18{]}; the size-three classification of Chapter 3 first
appeared in {[}23{]}. The monograph proves none of it. Its Chapter 1
names the gap directly - ``for the classification and efficient solving
of the stable marriage problem, characterization of the conditions for
decomposing instances of SMP to non-decomposable ones is critical.
Although several versions of decompositions of SMP are possible, their
characterization has not yet been completed'' - and its Section 3.5 adds
that the decomposition of an instance need not be unique. This paper
closes that gap for one precise sense of the word: a finest partition
into input-independent blocks always exists and is unique (Theorem 6),
the blocks are the square components of a filtration and so are
computable (Theorem 9.4), they form a partition exactly when the maximal
ones cover (Theorem 9.6), and the instances admitting none can be
counted at every size (Theorem 9). Apart from the sentence quoted above,
no text, figure, table or numerical value is taken from the monograph;
the figure below is generated from the definitions by the accompanying
code.

\emph{Recent work in other directions.} Deferred acceptance continues to
be studied from several directions that this paper does not touch:
agents who misreport or mistake their own preferences {[}24{]}, matroid
and partial-order constraints on what counts as a matching {[}25{]},
preferences that change after the fact {[}26,27{]}, distributed and
truthful implementations {[}28{]}, the running time of the algorithm
itself {[}29{]}, lattice-linear predicate methods for constrained
variants with ties {[}30{]}, and local search when ties are broken
{[}31{]}. Small instances have also been catalogued before, as integer
sequences {[}32{]} and in the OEIS {[}33{]}. The present paper is
orthogonal to all of these: it is about the structure of a single
instance and what that structure does to the algorithm run on it, and it
stays inside strict complete preferences throughout. Two further
preprints of the author study the same instance space from other angles
{[}34,35{]}.

\emph{How the statements are labelled.} Every numbered statement carries
a tag saying how it is established, because this paper mixes proofs with
complete computations and the difference matters. PROVED means a proof
is given here or cited. EXHAUSTIVE means the claim was checked on every
instance of the size named; at size three that is all 46,656 profiles
with strict complete preferences on both sides. EXACT means a finite
candidate set was fixed in advance by a proved statement and every
candidate in it was then settled, which is how sizes four to six are
reached where enumerating instances is out of the question. TARGETED
means a particular instance was constructed or searched for and is
supplied as a certificate. A tag never promotes a computation to a
proof: where a statement holds at every size the tag says PROVED, and
where it was verified only at certain sizes the tag names them.

\emph{Organisation.} Section 2 fixes the execution model and proves the
two facts the rest depends on: a reachable state is its pointer vector,
and the reachable set is scheduler-independent. Section 3 defines the
blocks, proves that a finest partition into them exists and is unique,
and characterises them as the square components of a filtration. Section
4 is the products theorem and the composition laws. Section 5 shows that
the converse fails and how far. Section 6 is the census, its asymptotics
and the recursion behind it. Section 7 lists five open problems. Two
companion papers are referred to: one on the Latin-rank-balance setting,
and a computational classification of the execution structure at small
sizes, in preparation, which is where the atlas of size-three executions
and the enumerable quotient at sizes five and six are developed.

\hypertarget{preliminaries-the-reachable-event-system}{%
\section{2. Preliminaries: the reachable event
system}\label{preliminaries-the-reachable-event-system}}

\hypertarget{model-and-reachable-event-system}{%
\subsubsection{Model and reachable event
system}\label{model-and-reachable-event-system}}

Let P = \{p\_0, \ldots, p\_(n-1)\} be proposers and R = \{r\_0, \ldots,
r\_(n-1)\} receivers. Unless stated otherwise every agent has a strict
complete order over the opposite side; such an instance is
\emph{balanced} and \emph{complete}. A state is s = (h, q), where h(r)
is the proposer currently held by r or the empty symbol, and q(p) is the
number of proposals p has already made. The canonical initial state has
h empty and q = 0. A proposer is \emph{active} at s when it is unheld
and still has an unproposed receiver. The local update U\_p advances
q(p) by one, sends the proposal to the next receiver on p's list, and
lets that receiver retain the better of its current holder and the new
proposer. A scheduler repeatedly picks an active proposer.

The event e = (p, i) denotes p making its (i+1)-st proposal. For a
pointer vector q the corresponding event set is F(q) = \{(p,i) : 0
\textless= i \textless{} q(p)\}. The rooted reachable digraph G(I)
contains every state reachable from the initial state under every legal
serial schedule; edges point away from the root and are coloured by
proposer. We also use the uncoloured digraph and the underlying
undirected graph.

A family of subsets of a finite event set is an \emph{antimatroid} when
it contains the empty set, is accessible, and is closed under union. A
\emph{poset antimatroid} is the family of down-sets of a partial order;
for the finite families here this is equivalent to being closed under
intersection as well. Intersection closure was tested separately from
the antimatroid axioms.

\hypertarget{two-notions-of-independence}{%
\subsubsection{Two notions of
independence}\label{two-notions-of-independence}}

\textbf{Input-independent partition (static).} Paired blocks P = P\_1 +
\ldots{} + P\_s and R = R\_1 + \ldots{} + R\_s with
\textbar P\_i\textbar{} = \textbar R\_i\textbar{} \textgreater{} 0, such
that every proposer in P\_i ranks all of R\_i above every receiver
outside R\_i, and every receiver in R\_i ranks all of P\_i above every
proposer outside P\_i. Equivalently, for each i the top
\textbar R\_i\textbar{} receivers of each p in P\_i are exactly R\_i,
and symmetrically on the other side. An instance is \emph{input-prime}
when no partition with s \textgreater= 2 exists. Candidate blocks were
generated from mutual top-k coalitions and assembled by exact cover,
because a mutual top coalition need not have an independent complement.

\textbf{Behavioural decomposition (dynamic).} Let E = E\_1 + \ldots{} +
E\_t partition the proposal events that actually occur, and set F\_i =
\{X intersect E\_i : X in F\}. The event system is \emph{behaviourally
decomposable} when F is the set of all disjoint unions X\_1 + \ldots{} +
X\_t with X\_i in F\_i. This preserves event identities and proposal
chains; it is strictly stronger than factoring the underlying uncoloured
undirected graph, and it is distinct from input independence.
Operationally, two proposers of one behavioural block are joined when
they propose to a common receiver during the run. One direction is
immediate (proposers that never share a receiver address disjoint
receiver sets and run independently), while the converse, that the
connected components of that graph are exactly the behavioural blocks,
is verified rather than proved; Section 4 states what was checked. The
graph is well defined because the event set is schedule-independent, and
the statements below rest on it only through the direction that is
proved, or at sizes where the two tests were both computed and agree.

\textbf{A third notion, and where the static one comes from.} A third
sense of the word is in the literature. Veklerov {[}18{]} calls an
instance decomposable when \emph{every} stable matching pairs a fixed
set of proposers with a fixed set of receivers and with no one else,
which is a condition on the stable set rather than on the preferences or
on the run. Input independence implies it, by Theorem 7, but behavioural
decomposition does not. Take n = 2 with p\_1 preferring r\_1 and p\_2
preferring r\_2, while r\_1 prefers p\_2 and r\_2 prefers p\_1. Each
proposer makes one proposal and is accepted, so the event system splits
into two single-event factors; but both perfect matchings are stable and
they disagree on both proposers, so no proper set of proposers is paired
with a fixed set of receivers. The third sense is implied by the static
notion and incomparable with the dynamic one. He gives three sufficient
criteria and one necessary one, and observes that a decomposition in his
sense need not be unique: an instance with a single stable matching has
every splitting of that matching as a decomposition. His first criterion
is the static condition of this paper, stated in the same words - a set
of proposers and an equinumerous set of receivers, each of which ranks
the other set above every outsider - and his second is a one-sided form
of it; {[}22, Chapter 7{]} restates both through the top-k operator and
attributes them to {[}18{]}. Neither is developed into a partition in
either place. The third sense does have a finest partition, and a unique
one: a partition is valid for it exactly when no edge of the union of
all stable matchings crosses it, so the valid partitions are the unions
of connected components of that union; each component is balanced,
because any stable matching restricts to a perfect matching on it, and
the component partition is therefore the finest. Both partitions are
functions of the input, and both can be computed without enumerating
anything exponential, since the pairs that lie in some stable matching
are found by the rotation machinery of {[}2{]}, which goes back to
Irving and Leather {[}36{]}. The difference is what the condition is
about. Veklerov's is a condition on the stable set, and it factors the
stable set, which is what it is about. The partition below is of the
other kind: the blocks are laminar, so a finest partition into them
exists, is unique (Theorem 6) and is determined by the ranked input
alone; they are the square components of a filtration (Theorem 9.4), so
the partition is computed from the preferences without the stable set;
they form a partition exactly when the maximal ones cover (Theorem 9.6);
and the instances admitting no such partition can be counted at every
size (Theorem 9). Theorem 8 separates the first two notions; the
relation of the third to the run is not pursued here.

Two earlier papers of the author's belong to this lineage, and naming
how they are weaker is the clearest way to say what is new. {[}37{]}
shows that when all proposers share one preference order the instance
reduces to n mutual-first pairs removed one at a time, so that its
stable set is the direct product of the stable sets of the pieces. That
is the singleton case of Theorem 7 for the stable set alone, and the
reduction is sequential: a pair is mutually first only after the earlier
pairs have been taken out, so it does not exhibit a partition of the
original input. The aligned family A\_n of Section 5 is not that
situation and is worth contrasting with it: there the proposers' first
choices are distinct and already mutual, so the singleton partition is
input-independent from the start, whereas under a shared order the first
choices coincide and a pair becomes mutually first only once the earlier
pairs are gone. The same paper calls two proposers k-indistinguishable
when their first k choices agree, observes that the classes refine as k
grows, and asks how deep into the lists one must look; the mutual-rank
filtration of Section 3 is the two-sided form of that question, and it
is mutuality rather than depth that makes the blocks laminar and the
finest partition unique. {[}38{]} takes the decomposition from the
blocking diagram instead, calls a top connected component
indecomposable, and observes that Veklerov's condition is its
proposer-oriented case. Both are one-sided, and both read the
decomposition off the stable set or the run rather than off the ranked
input.

The operator formulation of the run - a local update for each active
proposer, a scheduler choosing among them, and the progress invariants
that give termination, stability, proposer optimality and schedule
independence - is not reproved here. It is the subject of a companion
preprint and is cited as such. What is new below is the structure of the
FAMILY of reachable states, which that preprint does not study.

\hypertarget{lemma-1-states-are-determined-by-proposal-counts-proved}{%
\subsubsection{Lemma 1 (states are determined by proposal counts)
{[}PROVED{]}}\label{lemma-1-states-are-determined-by-proposal-counts-proved}}

\emph{At every reachable state the proposal-count vector q determines
the held-proposer vector h.}

\textbf{Proof.} For each proposer, q(p) fixes the initial segment of
receivers already contacted, so the multiset of past proposals is
reconstructed from q alone. Each receiver retains its most preferred
proposer among those it has received, so applying each receiver's order
to that reconstructed multiset yields h. No assumption of completeness
or balance is used. QED

The reachable graph can therefore be stored without a matching
coordinate, and q-injectivity holds: distinct reachable states have
distinct pointer vectors.

\hypertarget{theorem-2-local-commutativity-proved}{%
\subsubsection{Theorem 2 (local commutativity)
{[}PROVED{]}}\label{theorem-2-local-commutativity-proved}}

\emph{If two distinct proposers are active at a reachable state,
executing their next proposals in either order gives the same state
after both.}

\textbf{Proof.} Each pointer advances once under both orders. If the two
proposals target different receivers, the two holder updates are
disjoint. If they target the same receiver, the final holder is that
receiver's maximum among its previous holder and the two new proposers,
which does not depend on arrival order; the same agents are rejected
either way. QED

\hypertarget{lemma-3-persistent-enablement-proved}{%
\subsubsection{Lemma 3 (persistent enablement)
{[}PROVED{]}}\label{lemma-3-persistent-enablement-proved}}

\emph{If an event is enabled at s and a different proposer acts, the
event remains enabled until its own proposer executes it.}

\textbf{Proof.} An active proposer is unheld and has a next receiver on
its list. No other proposer's action can make it held, and no other
proposer consumes its next proposal. QED

\hypertarget{theorem-4-antimatroid-and-exchange-proved}{%
\subsubsection{Theorem 4 (antimatroid and exchange)
{[}PROVED{]}}\label{theorem-4-antimatroid-and-exchange-proved}}

\emph{The family F(I) of feasible proposal-prefix event sets is an
antimatroid, and any two complete executions with the same event set are
connected by a sequence of adjacent exchanges of events belonging to
different proposers.}

\textbf{Proof.} The empty set is feasible. Accessibility follows by
deleting the last event of any non-empty feasible prefix. For union
closure, take realisations of two feasible sets, start from the first,
and append the events present only in the second in their proposer-chain
order, skipping events already present; by Lemma 3 each required next
event is still enabled when reached. The same exchange argument moves a
chosen enabled event leftward through a complete execution, and
repetition connects any two complete executions with the same event set.
QED

\textbf{Corollary 4.1.} The feasible-set lattice is join-distributive.
Its join is set union, equivalently coordinatewise maximum of pointer
vectors. The meet need not be set intersection.

\textbf{Corollary 4.2.} F(I) is a poset antimatroid, and the lattice is
the distributive lattice of down-sets, exactly when F(I) is also
intersection-closed.

Intersection closure is common but not universal. EXHAUSTIVE at n = 3:
41,472 of the 46,656 profiles give poset antimatroids and 5,184 do not.
Of the 20,736 behaviourally prime profiles, 15,552 are poset and 5,184
are not; so behavioural primeness does not imply non-poset structure,
and every non-poset profile is behaviourally prime.

\textbf{Caution on trace theory.} The exchange statement of Theorem 4
resembles Mazurkiewicz trace equivalence, but the independence relation
here is state-dependent: two events are independent at some states and
not at others, because enablement depends on earlier rejections. The
system is therefore not a trace monoid over a fixed independence
alphabet; it is a concurrent transition system with the
strong-confluence property, and the attribution in the discussion is to
that weaker frame {[}11,12{]}. Local commutativity alone is not
sufficient for global confluence; the argument above uses persistent
enablement as well.

\hypertarget{theorem-5-boolean-cube-characterisation-proved}{%
\subsubsection{Theorem 5 (Boolean-cube characterisation)
{[}PROVED{]}}\label{theorem-5-boolean-cube-characterisation-proved}}

\emph{For a balanced instance with strict complete preferences, G(I) is
the n-dimensional Boolean cube if and only if the proposers' first
choices are pairwise distinct.}

\textbf{Proof.} If the first choices are distinct no proposal is ever
rejected, so each proposer has made either no proposal or exactly one,
every subset of first proposals is feasible, and the reachable graph is
the cube. Conversely, a first-choice collision in a complete balanced
market produces a rejected proposer that still has a proposal available,
hence a proposal chain longer than one coordinate. QED

\textbf{Corollary 5.1.} EXHAUSTIVE at n = 3: 10,368 of 46,656 profiles
are Boolean cubes. The count is also analytic: 48 of the 216 proposer
halves have distinct first choices, and the receiver half is
unconstrained, giving 48 x 216.

\textbf{The converse fails outside the hypotheses.} With incomplete
lists the cube can occur with a collision: two proposers whose lists are
the single receiver r\_0, with r\_0 ranking them in either order, give
the Boolean cube B\_2 although both first choices coincide. The
unbalanced instance with two proposers and one receiver behaves the same
way. Completeness and balance are therefore essential, not cosmetic.
{[}PROVED, with explicit certificates{]}

\clearpage

\hypertarget{blocks-and-the-unique-finest-partition}{%
\section{3. Blocks, and the unique finest
partition}\label{blocks-and-the-unique-finest-partition}}

\hypertarget{theorem-6-uniqueness-of-the-finest-input-partition-proved}{%
\subsubsection{Theorem 6 (uniqueness of the finest input partition)
{[}PROVED{]}}\label{theorem-6-uniqueness-of-the-finest-input-partition-proved}}

\emph{Every finite balanced instance with strict complete preferences
has a unique finest input-independent partition.}

\textbf{Proof.} Let (A,B) and (C,D) be blocks from two input-independent
partitions. Suppose A and C overlap and let p lie in the overlap. If
neither B nor D contains the other, pick b in B minus D and d in D minus
B. Membership of p in A forces b above d, and membership of p in C
forces d above b, contradicting strictness. Hence B and D are nested,
and the receiver-side argument nests A and C the same way. If A and C
are disjoint then so are B and D. The paired blocks of any two
partitions are therefore laminar, so the inclusion-minimal blocks of
their union form a common refinement that is again input-independent.
Iterating over the finite set of partitions gives a unique finest one.
QED

A useful consequence used repeatedly below: the finest partition of an
instance is the union of the finest partitions of its blocks, and a
block's restriction is input-prime exactly when that block cannot be
refined. The uniqueness is worth stating explicitly because the
literature records the decomposition of an instance as ambiguous.
Veklerov {[}18{]} exhibits an instance whose single stable matching
admits decompositions into two, three and four parts at once, and {[}22,
Section 3.5{]} repeats that the decomposition of an instance need not be
unique. That ambiguity is about decompositions in general and not about
finest ones, and taking the finest removes it in either sense of the
word, Veklerov's finest being the partition into connected components of
the union of the stable matchings. What Theorem 6 adds is that the
finest partition into input-independent blocks exists, is unique, and is
a function of the ranked input, so it is in hand before any stable
matching is computed.

Uniqueness on its own does not say how to find the partition, or how to
tell whether a given instance has one. Both follow from a graph
description of the blocks.

\hypertarget{the-structure-of-input-prime-instances}{%
\subsubsection{The structure of input-prime
instances}\label{the-structure-of-input-prime-instances}}

The blocks have a graph description that makes primality easy to see and
easy to test.

\textbf{The mutual-rank filtration.} For 1 \textless= k \textless= n let
M\_k be the bipartite graph on P together with R that joins p to r when
r is among the top k of p \emph{and} p is among the top k of r. These
nest,

M\_1 contained in M\_2 contained in \ldots{} contained in M\_n =
K\_(n,n),

and every vertex of M\_k has degree at most k. M\_1 is the graph of
mutual-first pairs, a partial matching; the filtration ends at the
complete bipartite graph. Call this the mutual-rank filtration of I. Its
one-sided ancestor is the k-indistinguishability of {[}37{]}, which
partitions the proposers alone by agreement of their first k choices;
the classes there also refine as k grows, but they pair with no
receivers, and it is the mutual edge that turns a nested family of
equivalence relations into blocks that are laminar on both sides.

\textbf{Theorem 9.4 (blocks are the square components of the
filtration).} {[}PROVED; EXHAUSTIVE at n = 3, verified at n = 4{]}
\emph{A pair (A, B) with \textbar A\textbar{} = \textbar B\textbar{} = k
is an input-independent block if and only if the union of A and B is a
connected component of M\_k isomorphic to K\_(k,k), equivalently a
component of M\_k in which every vertex has the maximum degree k.}

\textbf{Proof.} Let (A, B) be a block. Each p in A has top k equal to B,
and each r in B has top k equal to A, so every pair in A times B is an
edge of M\_k and the induced subgraph contains K\_(k,k). Since p has
only k receivers of rank at most k and all lie in B, p has no M\_k-edge
outside B, and symmetrically for receivers; so A together with B is a
full component and the induced graph is exactly K\_(k,k). Conversely, if
a component is a copy of K\_(k,k) on A and B with \textbar A\textbar{} =
\textbar B\textbar{} = k, then each p in A has k mutual edges, all into
B, and p has at most k receivers of rank at most k, so its top k is
exactly B; symmetrically on the other side. QED

\textbf{Proposition 9.5 (the blocks form a forest).} {[}PROVED{]}
\emph{Any two blocks are nested or disjoint on each side, so the blocks
of I, ordered by inclusion and rooted at the trivial block, form a
forest.}

\textbf{Proof.} This is the laminarity argument of Theorem 6, which
never used the assumption that the blocks belong to partitions: if p
lies in A intersect C with top \textbar B\textbar{} equal to B and top
\textbar D\textbar{} equal to D, then an element of B outside D and an
element of D outside B would have to be ranked above one another both
ways. QED

\textbf{Theorem 9.6 (primality is a covering failure).} {[}PROVED;
EXHAUSTIVE at n = 3{]} \emph{I is input-decomposable if and only if the
inclusion-maximal proper blocks cover P, equivalently cover R. Having a
proper block is not sufficient.}

\textbf{Proof.} The maximal proper blocks are pairwise disjoint by
Proposition 9.5, so if they cover they form a partition, and each is
input-independent by definition. Conversely any partition into proper
blocks consists of blocks contained in maximal ones, which therefore
cover. QED

The last sentence of the theorem is the point, and it is what makes
primality a subtle property rather than an obvious one. EXHAUSTIVE at n
= 3: 36,360 of the 46,656 instances, a little over seven in every nine,
possess a proper block and are nevertheless input-prime; the same holds
for 14,914 of 20,000 random size-four instances. The smallest witness
has n = 2 and one mutual-first pair: the pair is a block, the remaining
pair is not, and nothing covers, so a two-agent instance can carry a
block and stay prime. An input-prime instance is therefore not one
without local structure; it is one whose local structure fails to tile.

\textbf{Proposition 9.7 (the exact-cover polynomial, and why it
factorises).} {[}PROVED; EXHAUSTIVE at n = 3; TARGETED at n = 4 to 6{]}
\emph{Give every agent an arc to each of its top k choices on the other
side, and call a set of k proposers and k receivers a closed k-cell when
every such arc from a member of the set stays inside it. Then:}

\emph{(a) the closed k-cells are exactly the blocks of Theorem 9.4;}

\emph{(b) writing p(U) for the least proposer of a balanced pair (U, V),
the recursion H(empty, empty) = 1 and H(U, V) = the sum, over closed
cells (A, B) with p(U) in A inside U and B inside V, of z times H(U less
A, V less B) has the property that the coefficient of
z\textsuperscript{r} in H(P, R) is the number of exact covers of the
agents by r blocks; so I is input-prime exactly when H(P, R) = z;}

\emph{(c) H factorises over the block forest. If the maximal proper
blocks of a block b partition b, then H\_b = z + the product of H\_c
over those blocks; otherwise H\_b = z.}

\textbf{Proof.} (a) If (A, B) is a block of size k then every member of
A ranks exactly B in its first k places and every member of B ranks
exactly A, so no arc leaves the set. Conversely, closure of a balanced
set of 2k agents forces the k arcs of each member to exhaust the
opposite side of the set, which is the block condition. Closure
therefore delivers the mutual condition of Theorem 9.4 by itself, which
is why the directed form needs no reciprocity hypothesis and is the
shorter of the two definitions. (b) Distinguishing the least proposer
counts each cover once, because exactly one cell of a cover contains it,
and deleting that cell leaves an exact cover of the balanced remainder;
the converse adjoins one eligible cell to such a cover. (c) Blocks are
laminar by Proposition 9.5, so a block properly inside b lies inside one
of the maximal proper blocks of b, and those are pairwise disjoint. If
they fail to cover b, some agent of b lies in no proper block, so no
cover by proper cells exists and H\_b = z, which is Theorem 9.6. If they
do cover, every cover other than b itself restricts to an exact cover of
each of them, and any combination of such covers recombines into one of
b, which is the product. QED

Three things follow, and they are the reason for stating the polynomial
rather than only checking it. First, part (c) turns the recursion of
part (b) from a search over cells into a walk over the block forest,
linear in the size of that forest. Second, the degree of H is the number
of blocks in the finest partition and the leading coefficient is 1; that
coefficient is Theorem 6 read off the polynomial, since a second finest
factorisation would contribute a second monomial of top degree. Third, H
evaluated at 1, less 1, counts the non-trivial input-independent
partitions. One object therefore carries Theorem 6, Theorem 9.4 and
Theorem 9.6 at once, and this is the Hall-shaped recursive form that
Proposition 21 was reaching for: Hall's own induction separates a tight
set from the residual problem, and here the distinguished proposer
separates one closed cell from a balanced remainder.

\textbf{Table 17. The size-three census by exact-cover polynomial.}
{[}EXHAUSTIVE{]} Every one of the 46,656 profiles, classified by its
polynomial and by the two partitions the polynomial reconciles: the
coarsest non-trivial one, given by the maximal proper blocks, and the
finest one.

\begin{longtable}[]{@{}
  >{\raggedright\arraybackslash}p{(\columnwidth - 6\tabcolsep) * \real{0.2500}}
  >{\raggedright\arraybackslash}p{(\columnwidth - 6\tabcolsep) * \real{0.2500}}
  >{\raggedright\arraybackslash}p{(\columnwidth - 6\tabcolsep) * \real{0.2500}}
  >{\raggedright\arraybackslash}p{(\columnwidth - 6\tabcolsep) * \real{0.2500}}@{}}
\toprule\noalign{}
\begin{minipage}[b]{\linewidth}\raggedright
exact-cover polynomial
\end{minipage} & \begin{minipage}[b]{\linewidth}\raggedright
maximal proper blocks
\end{minipage} & \begin{minipage}[b]{\linewidth}\raggedright
finest partition
\end{minipage} & \begin{minipage}[b]{\linewidth}\raggedright
profiles
\end{minipage} \\
\midrule\noalign{}
\endhead
\bottomrule\noalign{}
\endlastfoot
z & none that cover & prime & 45,768 \\
z + z\textsuperscript{2} & 1+2 & 1+2 & 504 \\
z + z\textsuperscript{3} & 1+1+1 & 1+1+1 & 312 \\
z + z\textsuperscript{2} + z\textsuperscript{3} & 1+2 & 1+1+1 & 72 \\
\end{longtable}

The last row is the one to read. In 72 of the 888 decomposable
size-three instances the coarsest non-trivial partition and the finest
one differ: the two-agent block of the coarsening is itself decomposable
into singletons. The polynomial sees that and the shape of the maximal
blocks alone does not, which is worth keeping straight, because the
shape counts quoted elsewhere in this manuscript, 504 and 384, count
finest partitions; counted by maximal blocks the same profiles split 576
and 312. A second 72 sits nearby, the profiles with a non-trivial input
automorphism, and the two sets are disjoint; the coincidence of size is
only that, since an automorphism of order three cannot fix a partition
of shape 1+2.

\emph{Verification.} Over all 46,656 size-three profiles, and over 5,500
random and 9,500 planted decomposable profiles at n = 4, 5 and 6, six
things were checked and none failed: the closed cells agreed with the
blocks of Theorem 9.4; the coefficients of H agreed with a brute-force
enumeration of every exact cover; the criterion H = z agreed with
Theorem 9.6; the forest factorisation of part (c) agreed with the
recursion of part (b), coefficient by coefficient; the leading
coefficient was 1; and the degree was the number of blocks in the finest
partition. The coefficients are not all 0 or 1, which is the sign that
the object is doing work: a size-four instance with finest partition
1+1+1+1 and two disjoint two-agent blocks has polynomial z +
z\textsuperscript{2} + 2z\textsuperscript{3} + z\textsuperscript{4}, the
2 counting the two ways to split one of the pairs and not the other.

\clearpage

\hypertarget{one-partition-three-products}{%
\section{4. One partition, three
products}\label{one-partition-three-products}}

This is the section the paper exists for. The partition of Section 3 was
defined by a condition on the ranked input alone, and there is no reason
from the definition to expect it to say anything about the run or about
the stable set. It says everything about both.

\hypertarget{theorem-7-products-proved}{%
\subsubsection{Theorem 7 (products)
{[}PROVED{]}}\label{theorem-7-products-proved}}

\emph{An input-independent partition induces a Cartesian product of
rooted proposer-coloured execution graphs, a direct sum of
proposal-prefix antimatroids, and a blockwise product of stable
matchings and their lattices.}

\textbf{Proof.} No proposer ever reaches a receiver outside its block.
Suppose some proposer in P\_i made a first outside proposal; immediately
before it, every receiver in R\_i would have rejected it and would hold
some other proposer of P\_i, so \textbar R\_i\textbar{} = k receivers
would hold k distinct alternatives drawn from the k - 1 remaining
members of P\_i, which is impossible. Every update therefore changes one
block coordinate, which gives the graph product and the antimatroid
direct sum. Stability is block-diagonal, because a cross-block pair
would create a blocking pair with an internally preferred partner. QED

\textbf{Attribution.} The stable-matching part of Theorem 7 is not new.
For two blocks it is Definition 6.1 and Proposition 6.2 of {[}21{]},
where the same separation hypothesis gives Stab(S) = Stab(S1) x
Stab(S2), a product of lattices and a disjoint union of rotation posets.
What is new here is the extension to partitions with any number of
blocks, the uniqueness of the finest such partition (Theorem 6), and the
execution-graph product together with the antimatroid direct sum, none
of which concern the terminal stable set.

The three products are easiest to believe on one instance, and Fig 1 is
that instance: five proposers whose only proper blocks are the first
three and the last two, so that the finest partition has exactly two
parts and nothing finer exists. The digraph drawn there is at the same
time the Hasse diagram of the proposal-prefix antimatroid, so its being
a Cartesian product is the direct-sum statement as well, and the stable
set beside it is the product of the two factor stable sets. Every line
of the composition table below can be read off that one picture.

\clearpage

\begin{figure}[H]
\centering
\includegraphics[width=0.99\textwidth]{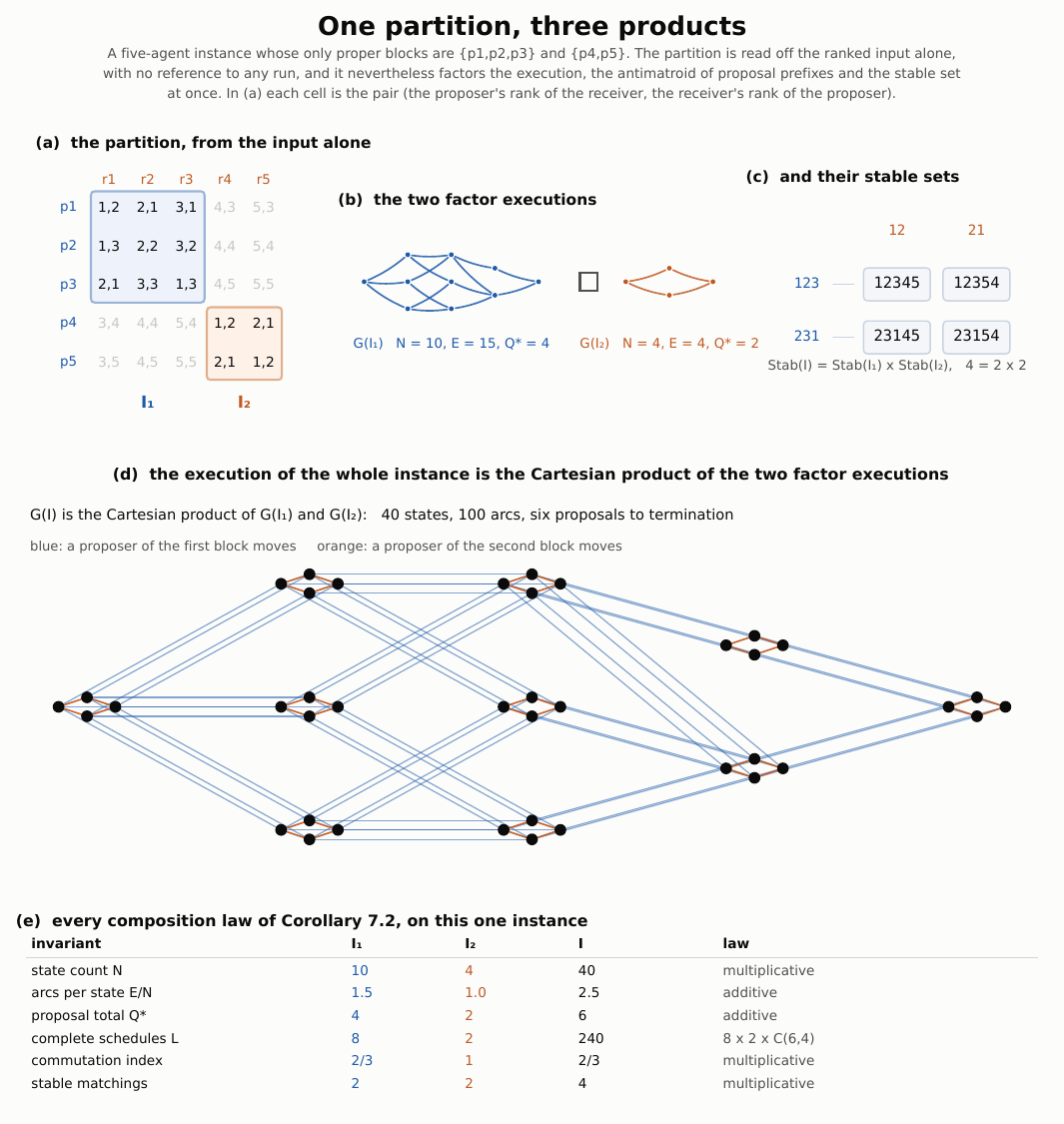}
\end{figure}

\textbf{Fig 1. One partition, three products.} {[}PROVED in general;
EXACT for the displayed instance{]} A five-agent instance I = I1 + I2
whose only proper blocks are \{p1,p2,p3\} and \{p4,p5\}, so the finest
input-independent partition has exactly two parts. Panel (a) is the
preference table, each cell carrying the pair (the proposer's rank of
the receiver, the receiver's rank of the proposer); the faded cells are
the cross-block pairs, which the proof of Theorem 7 shows no run ever
reaches. Panels (b) and (d) are the execution digraphs of the two
factors and of the whole instance, the latter drawn as ten copies of the
four-state factor, one at each state of the ten-state factor: blue arcs
move a proposer of the first block and orange arcs one of the second,
and the picture is the Cartesian product. Because that digraph is the
Hasse diagram of the proposal-prefix antimatroid, the same panel is the
direct-sum statement: the six events split 4 + 2 and a feasible prefix
is a pair of feasible prefixes. Panel (c) is the stable set, whose four
elements are exactly the pairs of a stable matching of each factor.
Panel (e) reads every line of Corollary 7.2 off this one instance: N =
10 x 4, E/N = 1.5 + 1.0, Q* = 4 + 2, L = 8 x 2 x C(6,4), the commutation
index 2/3 x 1, and the stable count 2 x 2. All six quantities are
computed from the definitions by the independent implementation, not
copied from the text.

\hypertarget{lemma-7.1-a-direct-sum-never-splits-a-proposer-proved-the-component-description-exhaustive-at-n-3}{%
\subsubsection{Lemma 7.1 (a direct sum never splits a proposer)
{[}PROVED; the component description EXHAUSTIVE at n =
3{]}}\label{lemma-7.1-a-direct-sum-never-splits-a-proposer-proved-the-component-description-exhaustive-at-n-3}}

\emph{In any direct-sum factorisation of the proposal-prefix
antimatroid, all events of one proposer lie in the same factor.
Consequently the behavioural factorisations are exactly the partitions
of the} proposer \emph{set for which the reachable pointer vectors form
the product of their projections.}

\emph{Proof.} Suppose two events of one proposer p lie in different
factors. The events of p are linearly ordered by the proposal that
produces them, and a proposal prefix contains an initial segment of
them, so if they do not all lie in one factor then some adjacent pair is
split: (p, i) in one factor and (p, i + 1) in another, with p making at
least i + 2 proposals. A feasible prefix stopping p after i proposals
meets the first factor in a set free of (p, i); a feasible prefix in
which p makes i + 2 meets the second factor in a set containing (p, i +
1). In a direct sum the union of those two pieces is feasible, but it
contains (p, i + 1) without (p, i) and is therefore not a proposal
prefix. So no factorisation separates two events of one proposer, and a
factorisation is determined by the induced partition of the proposers;
the direct-sum condition then says exactly that the set of reachable
pointer vectors is the product of its projections onto the parts.

The lemma matters twice. It makes the two decompositions of this paper
commensurable: both are now partitions of agents, the input one pairing
blocks of proposers with blocks of receivers and the behavioural one
partitioning the proposers alone, which is why Theorem 7 holds in the
direction it does and not the other. And it reduces the cost of deciding
behavioural primality from the 2\textsuperscript{Q*-1} bipartitions of
the event set to the 2\textsuperscript{n-1} - 1 partitions of the
proposers, which is what makes the size-four survey of a companion paper
in preparation possible at all.

\hypertarget{corollary-7.2-how-each-invariant-composes-and-why-it-matters-which-proved-exhaustive-at-n-3-sampled-at-n-4-to-6}{%
\subsubsection{Corollary 7.2 (how each invariant composes, and why it
matters which) {[}PROVED; EXHAUSTIVE at n = 3; SAMPLED at n = 4 to
6{]}}\label{corollary-7.2-how-each-invariant-composes-and-why-it-matters-which-proved-exhaustive-at-n-3-sampled-at-n-4-to-6}}

\emph{Over a direct sum the invariants of this paper compose in three
ways and no more. Table 16 is the whole content of that monoid picture,
read off one line at a time.}

\textbf{Table 16. Composition laws over a direct sum.} {[}PROVED{]} The
first five lines concern the behavioural direct sum of Lemma 7.1, a
partition of the proposers; the last concerns the input partition of
Theorem 7, which is the only one the stable set respects. The two-factor
forms are given here; the general r-factor forms are displayed a few
paragraphs below, and the column that matters is the last one. Verified
with 0 failures on 260 random direct sums of sizes four to six, and for
the stable set on all 888 decomposable size-three profiles.

\begin{longtable}[]{@{}
  >{\raggedright\arraybackslash}p{(\columnwidth - 4\tabcolsep) * \real{0.3333}}
  >{\raggedright\arraybackslash}p{(\columnwidth - 4\tabcolsep) * \real{0.3333}}
  >{\raggedright\arraybackslash}p{(\columnwidth - 4\tabcolsep) * \real{0.3333}}@{}}
\toprule\noalign{}
\begin{minipage}[b]{\linewidth}\raggedright
invariant
\end{minipage} & \begin{minipage}[b]{\linewidth}\raggedright
value on I + J
\end{minipage} & \begin{minipage}[b]{\linewidth}\raggedright
kind
\end{minipage} \\
\midrule\noalign{}
\endhead
\bottomrule\noalign{}
\endlastfoot
state count N & N(I) N(J) & multiplicative \\
arcs per state, E / N & E(I)/N(I) + E(J)/N(J) & additive \\
proposal total Q* & Q*(I) + Q*(J) & additive \\
complete schedules L & the multinomial of the block totals, times L(I)
L(J) & twisted \\
commutation index & the product of the two indices & multiplicative \\
stable-set size, over the input partition & the product of the two sizes
& multiplicative \\
\end{longtable}

\textbf{Proof.} The Cartesian product of Theorem 7 gives the state count
multiplicative and the arc count E(I)N(J) + E(J)N(I); dividing by
N(I)N(J) turns that into the additivity of arcs per state, which is the
form worth remembering. Every event belongs to exactly one factor, so
the proposal total is additive. A complete schedule of the sum is an
interleaving of one complete schedule of each factor, which gives the
multinomial for L. Dividing L by the multinomial of the full terminal
vector cancels the interleaving factor, so the commutation index is
multiplicative; that cancellation is the reason for normalising L this
way rather than any other. The last line is Theorem 7. QED

\emph{Remark (which invariants the algebra sees).} The reduced types -
the multisets of prime blocks - form a free commutative monoid under
direct sum, which is Theorem 6 read multiplicatively, so any invariant
that composes over the finest partition is fixed by its values on the
primes. Table 16 sorts those invariants into the two kinds the monoid
algebra distinguishes, and the sorting is what decides which of them can
be enumerated. The multiplicative ones are the evaluation homomorphisms
of the polynomial algebra; the additive ones are its gradings. A grading
has its candidate set fixed in advance by bounds on a single prime
factor, because the value on a direct sum is the sum over the parts and
bounds on the parts confine the total to an interval of integers. A
multiplicative invariant can be confined to a finite set the same way,
so it is not that gradings alone admit a finite candidate set; what the
grading gives is that the set is an interval, which is what the
certifiable quotient of a companion paper in preparation enumerates.
Confining a candidate set and settling every candidate in it remain
different things. Of the two gradings here only the proposal total is
integer-valued with both bounds proved, so it is the one that defines a
certifiable quotient and arcs per state is not. That is the reason that
quotient is the one it is.

The operational description given in Methods follows. The finest
behavioural partition is the partition of the proposers into connected
components of the graph joining p and p' when they propose to a common
receiver during the run. One direction is immediate: if no two proposers
on opposite sides of a partition ever address a common receiver, the two
groups touch disjoint receiver sets, run independently, and their
pointer vectors form a product. The converse we verified rather than
proved: it holds for every one of the 46,656 size-three profiles, and
for 40,000 random size-four and 8,000 random size-five profiles, with no
exception. The fast test was also checked against the event-set search
on all 46,656 size-three profiles, where the two agree everywhere. Call
an instance \emph{doubly prime} when it is prime both in the input sense
and in the behavioural sense. Behavioural primeness implies input
primeness, by the contrapositive of Theorem 7, so the doubly prime
instances are exactly the behaviourally prime ones; there are 20,736 of
them at n = 3.

With N\_i reachable states, E\_i edges, Q\_i terminal proposals and L\_i
complete schedules in factor i:

\[
N=\prod_i N_i,\qquad E=\sum_i E_i\prod_{j\neq i}N_j,\qquad Q^{*}=\sum_i Q_i^{*},\qquad L=\binom{Q^{*}}{Q_1^{*},\dots ,Q_r^{*}}\prod_i L_i .
\]

\hypertarget{the-commutation-index-proved-with-exhaustive-calibration}{%
\subsubsection{The commutation index {[}PROVED, with EXHAUSTIVE
calibration{]}}\label{the-commutation-index-proved-with-exhaustive-calibration}}

Define

\[
\kappa(I)=L(I)\Big/\binom{Q^{*}}{q^{*}_1,\dots ,q^{*}_n}.
\]

The denominator is the number of interleavings of the per-proposer
chains, that is, the number of complete executions an instance would
admit if all events of different proposers were independent everywhere.
So kappa measures how much the state-dependence of independence costs,
and kappa = 1 exactly when the system behaves like a free shuffle of its
chains. By the product formulas kappa is multiplicative over blocks, and
the Boolean cube has kappa = 1. Its value is a finer invariant than the
pair (N, E).

\clearpage

\hypertarget{the-converse-fails-and-how-badly}{%
\section{5. The converse fails, and how
badly}\label{the-converse-fails-and-how-badly}}

\hypertarget{theorem-8-the-converse-fails-proved}{%
\subsubsection{Theorem 8 (the converse fails)
{[}PROVED{]}}\label{theorem-8-the-converse-fails-proved}}

\emph{Behavioural decomposability does not imply input decomposability.}

\textbf{Proof.} Take any instance whose proposers have pairwise distinct
first choices. By Theorem 5 the execution graph is the Boolean cube,
whose event system is the direct sum of n one-event systems, so the
instance is behaviourally decomposable. The receiver orders are
unconstrained, and choosing them so that no mutual top-k coalition is
complemented destroys every input-independent partition. QED

EXHAUSTIVE at n = 3: 25,920 profiles are behaviourally decomposable
while only 888 are input-decomposable, so 25,032 behaviourally
decomposable profiles are input-prime. Conversely every behavioural
block lies inside an input block, with no exceptions among the 46,656
profiles. Proposition 12.3 turns this into a named family valid at every
size: the cyclic anti-phase profile C\_n is input-prime while its
execution is the Boolean cube, whose event system is the direct sum of n
one-event systems.

The behavioural side has a uniqueness theorem of its own, and comparing
the two is what shows how far apart they are.

\hypertarget{theorem-6.1-the-behavioural-factorisation-is-unique-too-proved-exhaustive}{%
\subsubsection{Theorem 6.1 (the behavioural factorisation is unique too)
{[}PROVED;
EXHAUSTIVE{]}}\label{theorem-6.1-the-behavioural-factorisation-is-unique-too-proved-exhaustive}}

\emph{Let F be any family of subsets of a finite set E. The partitions
of E over which F factorises as a direct sum are closed under common
refinement. Consequently F has a unique finest direct-sum factorisation,
namely the meet of all of them.}

\textbf{Proof.} F is always contained in the product of its projections,
so F factorises over a partition exactly when the size of F equals the
product of the projection sizes; in particular the one-block partition
is always valid. Let A and B be valid partitions, and for every C in A
and D in B choose an element x(C,D) of the projection of F onto C
intersect D. Fix C. For each D pick a witness X(C,D) in F whose
restriction to C intersect D is x(C,D). Because B is valid, assembling
the D-parts of these witnesses gives an element U of F with U restricted
to D equal to X(C,D) restricted to D for every D; hence U restricted to
C is the disjoint union over D of the x(C,D), and that union lies in the
projection of F onto C. Because A is valid, assembling these C-parts
gives the disjoint union over all C and D of the x(C,D) as an element of
F. So F factorises over the common refinement of A and B. The meet of
the finitely many valid partitions is therefore itself valid, and it
refines every valid partition. QED

\textbf{Corollary 6.2 (unique prime decomposition).} Both decompositions
used in this paper are unique. On the input side the finest partition is
unique by Theorem 6 and its blocks are input-prime; on the behavioural
side the finest direct-sum factorisation of the proposal-prefix family
is unique by Theorem 6.1 and its blocks are behaviourally prime. In
either case, if an instance is written as a composition of factors that
are all prime, that decomposition is the finest one, so the factors are
determined by the instance - not merely up to isomorphism, but as a
partition of the labelled agent set. No ordering convention is needed.

EXHAUSTIVE support. Over all 255 non-empty set families on three events
and all 65,535 on four events, no pair of valid partitions has an
invalid meet and every family has a valid finest factorisation. Over all
46,656 size-three instances the same holds for the proposal-prefix
family, and the finest factorisation never splits one proposer's events
between blocks. Its block sizes are 1+1+1 for the 10,368 Boolean cubes,
1+3 for 15,552 instances, and a single block of 5, 6 or 7 events for
14,256, 5,832 and 648 instances respectively - which is the event-level
form of the behavioural counts.

\textbf{Remark 6.3 (commutativity is not the mechanism).} It is tempting
to argue that, because composition of independent blocks is commutative
and associative, uniqueness follows from the definition of a prime
instance alone. It does not. Commutativity and associativity together
with irreducibility are not sufficient for unique factorisation: in the
numerical semigroup generated by 2 and 3 under addition, both generators
are irreducible, yet 6 = 2 + 2 + 2 = 3 + 3 has two factorisations into
irreducibles, of different lengths. What is needed is the stronger
property that the valid decompositions are closed under common
refinement - Theorem 6 on the input side, Theorem 6.1 on the behavioural
side. The blocks here are prime in that strong sense, not merely
irreducible, and it is the strong sense that the proofs use.

\textbf{Remark 6.4 (an imposed order is a normal form; a derived order
can be more).} Uniqueness also cannot be obtained by \emph{imposing} an
order on the factors of a commutative decomposition. Such an ordering
replaces a multiset of factors by a sequence, which removes the
ambiguity of which factor is written first but cannot identify two
genuinely different factorisations: sorting the two factorisations of 6
above still leaves (2,2,2) and (3,3). Where factors are sorted in this
paper - by block size and then by the canonical form of the sub-instance
- that is only to obtain a canonical sequence for hashing and for the
certificates in the supplementary material. An order \emph{derived from
the structure} rather than imposed on it is a different matter, and the
execution space carries exactly such an order.

\textbf{Remark 6.5 (where a non-commutative composition would enter).} A
non-commutative composition exists nearby and is not covered by these
theorems. Require only the proposer half of the block condition: every
proposer of P\_1 ranks all of R\_1 above everything outside R\_1, with
no condition on the receivers of R\_1. This relation is not symmetric in
the two blocks, and it admits no product theorem. Witness: the instance
in which all six agents rank the opposite side in the identical order.
Proposer p\_1 has top choice r\_0, so the pair (\{p\_1\},\{r\_0\})
satisfies the proposer half; but r\_0 ranks p\_0 first, the instance is
behaviourally prime, and its execution has 15 states and 6 terminal
proposals, so nothing factors. The one-sided condition is also nearly
vacuous as a decomposition criterion: every instance satisfies it for
some singleton pair outright, since any proposer paired with its own
first choice will do, and 46,272 of the 46,656 size-three instances
satisfy it for some singleton pair that is not itself a block, including
all 20,736 behaviourally prime ones; the remaining 384 are those in
which every proposer is mutually first-matched. Both directions of
dominance are therefore needed, and it is the symmetric condition that
makes the composition commutative in the first place.

\textbf{Proposition 9.10 (the dynamic decomposition structure is flat,
and its counting degenerates).} {[}PROVED; EXHAUSTIVE at n = 3; EXACT at
n = 4 to 6{]} \emph{(a) The behavioural factorisations of an instance
are exactly the coarsenings of the unique finest one. So if the finest
has r parts the instance has exactly B(r) of them, B the Bell number,
and that count depends on nothing else about the instance.}

\emph{(b) By Lemma 7.1 the finest factorisation is a partition of the
proposers, so the dynamic decomposition structure is a partition of n
and not a forest: there is no nesting to record.}

\emph{(c) Every partition of n is the shape of the finest factorisation
of some input-prime instance at n = 3, 4, 5 and 6, so at those sizes the
dynamic decomposition-level classification is complete, with exactly
p(n) classes: 3, 5, 7 and 11.}

\textbf{Proof.} (a) Theorem 6.1 makes the factorisations closed under
common refinement, so a finest one exists and is unique. If s is any
factorisation, the meet of s with the finest is a factorisation at or
below the finest, hence equal to it, so the finest refines s. Conversely
merging factors of a direct sum leaves a direct sum, so every coarsening
of the finest is a factorisation. The factorisations are therefore in
bijection with the set partitions of the r parts, and there are B(r) of
those. (b) is Lemma 7.1. (c) by exhibiting an instance for each
partition. QED

\emph{Verification.} Part (a) was checked by brute force, enumerating
every partition of the proposers and testing the product condition
directly, over all 46,656 size-three profiles and 4,800 random size-four
and size-five ones. The count was B(r) every time, the pairs of the
number of prime factors and the number of factorisations taking only the
values (1, 1), (2, 2), (3, 5), (4, 15) and (5, 52).

The two families below are the sharpest separation this paper has, and
the symmetry needed to describe them is short enough to give here.

An \emph{instance automorphism} is a pair g = (g\_P, g\_R) of
permutations of the two sides that carries the ranked bipartite graph to
itself. Aut(I) is defined as a subgroup of S\_n x S\_n and deliberately
excludes the exchange of the two sides; side exchange is treated
separately below.

\hypertarget{lemma-10-rigidity---recalled-from-21-lemma-4.1}{%
\subsubsection{Lemma 10 (rigidity) - recalled from {[}21, Lemma
4.1{]}}\label{lemma-10-rigidity---recalled-from-21-lemma-4.1}}

\emph{If g in Aut(I) fixes one agent then g is the identity.}

\textbf{Proof.} Suppose g\_P(p\_0) = p\_0. Preservation of p\_0's strict
complete order over all receivers forces g\_R to be the identity. With
g\_R trivial, preservation of every receiver's strict complete order
over all proposers forces g\_P to be the identity. The same argument
runs with the sides interchanged. Completeness of the lists is used in
both steps. QED

\hypertarget{proposition-11-free-action-and-divisibility---recalled-from-21-proposition-4.2-semiregularity-from-39}{%
\subsubsection{Proposition 11 (free action and divisibility) - recalled
from {[}21, Proposition 4.2{]}, semiregularity from
{[}39{]}}\label{proposition-11-free-action-and-divisibility---recalled-from-21-proposition-4.2-semiregularity-from-39}}

\emph{Aut(I) acts freely on P and on R. Consequently
\textbar Aut(I)\textbar{} divides n, every non-identity automorphism is
fixed-point-free on both sides, and every orbit has size
\textbar Aut(I)\textbar.}

Lemma 10 and Proposition 11 are stated for self-containedness and are
not claimed as contributions. They are Lemma 4.1 and Proposition 4.2 of
{[}21{]}, which itself identifies Proposition 11 as the profile-level
specialisation of a semiregularity property of preference-profile
stabilisers proved in {[}39{]}. Following {[}21, Definition 4.4{]} we
call a profile automorphism-maximal when its automorphism group has
order n, that is when Aut(I) acts regularly on each side; {[}21, Theorem
4.5{]} puts every such profile in a regular normal form over a group of
order n.

Everything this paper derives from symmetry flows from Proposition 11.
For odd n a side-preserving involution is impossible. For prime n = p,
Aut(I) is trivial or cyclic of order p acting regularly, so non-trivial
and automorphism-maximal coincide at prime size. The coincidence does
not survive to n = 4, where a group of order two is non-trivial without
being maximal. Non-abelian automorphism groups first become possible at
n = 6, since a group acting freely on at most five points has order 1,
2, 3, 4 or 5 and is abelian. What is new below is none of this, but the
passage from these constraints to the execution digraph, and the exact
counting of symmetric instances.

\hypertarget{theorem-12-regular-symmetry-forces-the-boolean-cube-proved}{%
\subsubsection{Theorem 12 (regular symmetry forces the Boolean cube)
{[}PROVED{]}}\label{theorem-12-regular-symmetry-forces-the-boolean-cube-proved}}

\emph{If \textbar Aut(I)\textbar{} = n, so that Aut(I) acts regularly on
each side, then G(I) is the n-dimensional Boolean cube.}

\textbf{Proof.} Write f(p) for the first choice of p.~Equivariance gives
f(g\_P(p)) = g\_R(f(p)). Since the action on P is simply transitive, f
is determined by its value at one proposer, and since the action on R is
simply transitive, f is a bijection. The first choices are therefore
pairwise distinct, and Theorem 5 applies. QED

The converse is false: Boolean-cube execution does not require any input
symmetry at all, as the next subsection quantifies.

\hypertarget{corollary-12.1-the-regular-anti-phase-templates-run-as-a-boolean-cube-proved-exhaustive-for-orders-at-most-5}{%
\subsubsection{Corollary 12.1 (the regular anti-phase templates run as a
Boolean cube) {[}PROVED; EXHAUSTIVE for orders at most
5{]}}\label{corollary-12.1-the-regular-anti-phase-templates-run-as-a-boolean-cube-proved-exhaustive-for-orders-at-most-5}}

\emph{Every automorphism-maximal profile has Boolean-cube execution. In
particular so does every regular anti-phase template P(G, A) of
{[}21{]}, for every finite group G and every ordering A.}

\textbf{Proof.} In the regular normal form P(G, A, B) of {[}21, Theorem
4.5{]} the first choice of the proposer indexed by g is the receiver
indexed by g a\_0, and g maps to g a\_0 is a bijection of G. First
choices are therefore pairwise distinct and Theorem 5 applies. QED

So on any such instance deferred acceptance makes exactly n proposals,
none of which is rejected, and stops at the matching g maps to g a\_0.
The execution invariants are forced: N = 2\textsuperscript{n}, E = n
2\textsuperscript{n-1}, Q* = n, L = n! and kappa = 1.

\textbf{Table 4a. Regular anti-phase templates, execution against stable
set.} Computed directly; the stable-set column agrees with {[}21,
Observation 7.2{]}.

\begin{longtable}[]{@{}llllllll@{}}
\toprule\noalign{}
G & normalised orderings & N & E & Q* & L & kappa & size of Stab \\
\midrule\noalign{}
\endhead
\bottomrule\noalign{}
\endlastfoot
Z2 & 1 & 4 & 4 & 2 & 2 & 1 & 2 \\
Z3 & 2 & 8 & 12 & 3 & 6 & 1 & 3 \\
Z4 & 6 & 16 & 32 & 4 & 24 & 1 & 4, 6 or 8 \\
V4 & 6 & 16 & 32 & 4 & 24 & 1 & 10 \\
Z5 & 24 & 32 & 80 & 5 & 120 & 1 & 5 \\
\end{longtable}

\hypertarget{proposition-12.2-execution-structure-and-the-stable-set-are-independent-invariants-proved-with-certificates-exhaustive-at-n-3}{%
\subsubsection{Proposition 12.2 (execution structure and the stable set
are independent invariants) {[}PROVED, with certificates; EXHAUSTIVE at
n =
3{]}}\label{proposition-12.2-execution-structure-and-the-stable-set-are-independent-invariants-proved-with-certificates-exhaustive-at-n-3}}

\emph{The proposer-coloured execution digraph does not determine the
number of stable matchings, and the number of stable matchings does not
determine the execution digraph.}

\textbf{Certificates.} All six normalised orderings of the cyclic group
of order four give templates with the identical execution digraph of
Table 4a, while their stable sets have 4, 6 and 8 elements {[}21,
Observation 7.2{]}; this proves the first half. The Klein template has
the same execution digraph again and ten stable matchings, the classical
maximum f(4) at size four {[}21, Observation 5.13{]}, the maximum itself
being known to grow simply exponentially {[}13{]}; so the size-four
instance with the largest stable set has the smallest execution graph a
size-four instance can have. For the second half, an exhaustive pass
over all 46,656 size-three profiles shows that every value of the
stable-matching count occurs in several distinct execution classes.

\hypertarget{proposition-12.3-two-latin-families-that-the-execution-cannot-tell-apart-proved-exhaustive-for-n-at-most-6}{%
\subsubsection{Proposition 12.3 (two Latin families that the execution
cannot tell apart) {[}PROVED; EXHAUSTIVE for n at most
6{]}}\label{proposition-12.3-two-latin-families-that-the-execution-cannot-tell-apart-proved-exhaustive-for-n-at-most-6}}

The sharpest form of the separation uses two families that exist at
every size. Both are rank-Latin and automorphism-maximal, so by
Corollary 12.1 both run as the Boolean cube; they sit at opposite
extremes of everything the execution digraph cannot see. A strict-Latin
instance has no proper block at all, at any size, and for a short
reason: a proposer's first choice lies in its own block, and the Latin
condition makes that receiver rank the proposer n-th, so a block
containing the pair must have size at least n.~It is prime in the third
sense too, and for the reason that makes the Latin instances interesting
in the first place: for each t the matching pairing every proposer with
its t-th choice is stable, because a blocking pair would need ranks
summing to less than n + 1. That is the stability criterion of Benjamin,
Converse and Krieger discussed below, and the author restated it as
{[}37, Theorem 2{]}. The n rank-matchings therefore exhaust K\_(n,n)
between them, the union of the stable matchings is complete, and the
finest partition in the third sense is the trivial one. C\_n is
therefore prime in the input sense and in the third sense at once, and
at the same time it runs as the Boolean cube, which is the most
decomposable behaviour there is: its n proposers make one proposal each,
to n distinct receivers, so the event system splits into n single-event
factors. That is the separation of Theorem 8 in its sharpest form. The
statement is about C\_n and the strict-Latin condition; A\_n is
rank-Latin but not strict-Latin, and sits at the opposite end of the
input axis. The four-agent Latin instance of {[}38, Example 8{]} is
decomposed there in the blocking-diagram sense of that paper, which is a
condition on neither the input nor the stable set and is not compared
with either here. Latin instances in the strict sense, where the rank of
a receiver in a proposer's list and of that proposer in the receiver's
list sum to n + 1, are the latin marriages of Benjamin, Converse and
Krieger {[}14{]} and go back to Knuth {[}19{]}. Their stability
criterion forces at least n stable matchings on any such instance, and
their family S\_n attains that minimum; C\_n is S\_n up to relabelling,
which is a second derivation of the count in the table. A\_n has a
single stable matching at every size and so cannot be Latin in that
sense at all; it is rank-Latin only in the weaker sense used here, where
each rank level is a perfect matching on each side. {[}EXHAUSTIVE for n
= 2 to 6{]}

The \emph{aligned cyclic Latin profile} A\_n gives the receiver indexed
by j rank 1 + ((j - i) mod n) in the list of the proposer indexed by i,
and symmetrically on the other side, so every pair (p\_i, r\_i) is
mutual first. It is the profile introduced in the companion submission
on Latin rank balance. The \emph{cyclic anti-phase profile} C\_n has the
proposer indexed by i rank the receivers from i upwards and the receiver
indexed by j rank the proposers from j+1 upwards; it is the profile of
{[}21, Section 3{]}, and it has no mutual-first pair at all.

\emph{For every n at least 2, A\_n and C\_n have the same reachable
execution digraph, namely the Boolean cube with N =
2\textsuperscript{n}, E = n 2\textsuperscript{n-1}, Q} = n, L = n! and
kappa = 1, and both have automorphism group of order n.~Their finest
input partitions are the all-singletons partition and the one-block
partition respectively, so A\_n is maximally input-decomposable and C\_n
is input-prime; and their stable sets have 1 and n elements
respectively.*

\textbf{Table 4d. The two cyclic Latin families.} {[}EXHAUSTIVE for n =
2 to 6; the execution row follows from Corollary 12.1 for every n{]}

\begin{longtable}[]{@{}lll@{}}
\toprule\noalign{}
& A\_n, aligned & C\_n, anti-phase \\
\midrule\noalign{}
\endhead
\bottomrule\noalign{}
\endlastfoot
rank-Latin & yes & yes \\
order of Aut & n & n \\
execution digraph & Boolean cube B\_n & Boolean cube B\_n \\
commutation index kappa & 1 & 1 \\
mutual-first pairs & n, a perfect matching & 0 \\
finest input partition & 1+1+\ldots+1 & one block, input-prime \\
size of the stable set & 1 & n \\
\end{longtable}

\textbf{Proof.} Both profiles are invariant under simultaneous rotation
of the two sides, so each has automorphism group of order at least n,
and Proposition 11 forces equality; Corollary 12.1 then gives the
Boolean cube and its invariants. In A\_n the pair (p\_i, r\_i) ranks
each other first for every i, so the all-singletons partition is
input-independent and, being the finest possible, is the finest; the
stable set is a single matching because a perfect matching of
mutual-first pairs is forced. In C\_n the top choice of the proposer
indexed by i is the receiver indexed by i, whose own top choice is the
proposer indexed by i+1, so no block of size one exists; for 1
\textless{} k \textless{} n the top k receivers of the proposer indexed
by i are the interval from i to i+k-1, which moves with i, so no block
of size k exists either and C\_n is input-prime. Its stable set has n
elements {[}21, Proposition 3.3{]}. QED

This one family settles three things at once and at every size, where
the earlier certificates settled them only at n = 4. The execution
digraph does not determine the stable-matching count, since A\_n and
C\_n share it and differ by a factor of n.~It does not determine input
decomposability either, which sharpens Theorem 8 from an existence
statement to a named family: C\_n is input-prime with fully parallel
behaviour. And rank-Latinness, a strong input regularity, constrains
none of it - the point the companion submission makes for the stable
set, holding here for the execution as well.

\clearpage

\hypertarget{the-exact-census}{%
\section{6. The exact census}\label{the-exact-census}}

\hypertarget{theorem-9-exact-census-of-input-prime-instances-proved-exact}{%
\subsubsection{Theorem 9 (exact census of input-prime instances)
{[}PROVED;
EXACT{]}}\label{theorem-9-exact-census-of-input-prime-instances-proved-exact}}

Let T(n) = (n!)\textsuperscript{2n} be the number of balanced strict
complete instances of size n, let Pi(n) be the number of input-prime
ones, and for a partition lambda = (k\_1, \ldots, k\_s) of n with
multiplicities m\_j write N\_lambda(n) for the number of paired block
structures of that shape. Then Pi(1) = 1 and

\[
T(n)=\sum_{\lambda=(k_1,\dots ,k_s)\vdash n} N_\lambda(n)\prod_{i=1}^{s}\Pi(k_i)\big((n-k_i)!\big)^{2k_i},\qquad N_\lambda(n)=\frac{1}{\prod_{j\ge 1}m_j(\lambda)!}\left(\frac{n!}{\prod_i k_i!}\right)^{2}.
\]

\emph{The lambda-term counts exactly the instances whose finest
partition has shape lambda, and the s = 1 term is Pi(n), so the identity
determines Pi(n) recursively.}

\textbf{Proof.} By Theorem 6 the finest partition is unique, so the
instances are partitioned by its shape. Fix a block structure with
blocks (P\_i, R\_i) of sizes k\_i. A proposer in P\_i has a list
consisting of an arbitrary order of R\_i followed by an arbitrary order
of its complement: k\_i! times (n - k\_i)! choices, of which the first
factor is exactly the datum of the restricted sub-instance and the
second is free. The same holds on the receiver side. The structure is
the finest partition precisely when each restricted sub-instance is
input-prime, since the finest partition of the whole instance is the
union of the finest partitions of its blocks. Collecting the free tails
gives ((n - k\_i)!)\textsuperscript{2k\_i} per block and the restricted
data gives Pi(k\_i). Counting the block structures of shape lambda gives
N\_lambda(n). QED

\textbf{Table 3. Exact census of input-prime instances.}

\begin{longtable}[]{@{}
  >{\raggedright\arraybackslash}p{(\columnwidth - 8\tabcolsep) * \real{0.2000}}
  >{\raggedright\arraybackslash}p{(\columnwidth - 8\tabcolsep) * \real{0.2000}}
  >{\raggedright\arraybackslash}p{(\columnwidth - 8\tabcolsep) * \real{0.2000}}
  >{\raggedright\arraybackslash}p{(\columnwidth - 8\tabcolsep) * \real{0.2000}}
  >{\raggedright\arraybackslash}p{(\columnwidth - 8\tabcolsep) * \real{0.2000}}@{}}
\toprule\noalign{}
\begin{minipage}[b]{\linewidth}\raggedright
n
\end{minipage} & \begin{minipage}[b]{\linewidth}\raggedright
instances T(n)
\end{minipage} & \begin{minipage}[b]{\linewidth}\raggedright
input-decomposable
\end{minipage} & \begin{minipage}[b]{\linewidth}\raggedright
input-prime Pi(n)
\end{minipage} & \begin{minipage}[b]{\linewidth}\raggedright
decomposable fraction
\end{minipage} \\
\midrule\noalign{}
\endhead
\bottomrule\noalign{}
\endlastfoot
1 & 1 & 0 & 1 & 0 \\
2 & 16 & 2 & 14 & 1.250e-1 \\
3 & 46,656 & 888 & 45,768 & 1.903e-2 \\
4 & 110,075,314,176 & 88,478,208 & 109,986,835,968 & 8.038e-4 \\
5 & 6.1917e20 & 11,557,648,579,461,120 & 6.1916e20 & 1.867e-5 \\
6 & 1.9408e34 & 7.8739e27 & 1.9408e34 & 4.057e-7 \\
7 & 6.8238e51 & 5.7129e43 & 6.8238e51 & 8.372e-9 \\
8 & 4.8790e73 & 7.5836e63 & 4.8790e73 & 1.554e-10 \\
\end{longtable}

The n = 3 row reproduces the exhaustive enumeration, and the n = 4 row
was confirmed independently by the side-splitting enumeration described
in Methods. At n = 4 the finest partition has shape 3+1 for 26,362,368
profiles, 2+2 for 903,168, 2+1+1 for 20,901,888 and 1+1+1+1 for
40,310,784. The size-five row is exact too: T(5) =
619,173,642,240,000,000,000 and Pi(5) = 619,162,084,591,420,538,880,
abbreviated in the table only to keep the column narrow.

\textbf{Remark 9.3 (blocks of size one are mutual-first pairs).} An
input-independent block of size one is a pair that ranks each other
first. Such pairs are the \emph{mutual-first pairs} of the companion
submission on Latin rank balance, where they are shown to be pairwise
disjoint and to belong to every stable matching, so that a perfect
matching of them certifies a unique stable matching. In the language
used here that is the extreme case of Theorem 7: when the finest
partition is all singletons, every factor carries a single proposal
event, the execution graph is the Boolean cube, and the stable-matching
lattice is a product of one-element lattices. Theorem 9 counts those
instances exactly. The all-singletons term is n! times
((n-1)!)\textsuperscript{2n} - that is 2 at n = 2, 384 at n = 3 and
40,310,784 at n = 4 - so the forced-pair criterion applies to a fraction
n!/n\textsuperscript{2n} of all instances, and by Corollary 9.1 that
single term dominates input decomposability as n grows.

What the present framework adds is the case k at least 2. A mutual-first
pair is a mutual top-one coalition; an input block of size k is a mutual
top-k coalition. Theorem 6 shows that these are laminar for all k
simultaneously, which is why a finest partition exists and is unique,
and that in turn is what makes the product formulas of Theorem 7 and the
census of Theorem 9 possible. None of it follows from the size-one case,
and conversely nothing here supplies the certificate theory for the
stable set that the companion submission develops.

\textbf{Corollary 9.1 (asymptotics, with the leading correction).}
{[}PROVED; EXACT at n = 2 to 12{]} \emph{Write W(lambda) for the
lambda-term of the census identity. Then W(1\textsuperscript{n}) is
exactly the fraction n!/n\textsuperscript{2n} of T(n); the shape (2,
1\textsuperscript{n-2}) contributes exactly
7n/(2(n-1)\textsuperscript{3}) of W(1\textsuperscript{n}), and the shape
(2, 2, 1\textsuperscript{n-4}) exactly
49n(n-2)(n-3)/(8(n-1)\textsuperscript{7}); and all remaining shapes with
at least two parts together contribute O(n\textsuperscript{-6}) of it.
The one-part shape lambda = (n) is the prime term Pi(n) itself and is
not part of D(n), so the quantifier here and the bound below range over
the shapes with every part at most n - 1. Hence}

D(n)/T(n) = (n!/n\textsuperscript{2n}) (1 +
7n/(2(n-1)\textsuperscript{3}) + O(n\textsuperscript{-4})) =
(n!/n\textsuperscript{2n})(1 + Theta(n\textsuperscript{-2})).

\textbf{Proof.} For lambda = 1\textsuperscript{n} the multiplicity
divisor is n!, so N = (n!)\textsuperscript{2}/n! = n!, and each block
contributes Pi(1)((n-1)!)\textsuperscript{2}; hence
W(1\textsuperscript{n}) = n!((n-1)!)\textsuperscript{2n}, which against
T(n) = (n!)\textsuperscript{2n} is n!/n\textsuperscript{2n}. For lambda
= (2, 1\textsuperscript{n-2}) the divisor is (n-2)!, giving N =
(n!)\textsuperscript{2}/(4(n-2)!), while the blocks contribute
Pi(2)((n-2)!)\textsuperscript{4} = 14((n-2)!)\textsuperscript{4} and
((n-1)!)\textsuperscript{2(n-2)}; dividing by W(1\textsuperscript{n})
and using (n-1)! = (n-1)(n-2)! and n! = n(n-1)(n-2)! leaves
7n/(2(n-1)\textsuperscript{3}), the 7/2 being Pi(2)/4. The same
computation with two blocks of size two, whose divisor carries an extra
2!, gives 49n(n-2)(n-3)/(8(n-1)\textsuperscript{7}).

For the remaining shapes bound Pi(k) by T(k) = (k!)\textsuperscript{2k}
and put g(k) = k!(n-k)!/(n-1)!, which is k/C(n-1,k-1), so that g(1) = 1
and g(k) \textless= 1 throughout 1 \textless= k \textless= n-1.
Replacing the falling factorial n!/(n-K)! by n\textsuperscript{K}, where
K is the number of agents lying in blocks of size at least two, gives

W(lambda)/W(1\textsuperscript{n}) \textless= (1/prod\_(j\textgreater=2)
m\_j!) prod\_(k\_i \textgreater= 2) phi(k\_i), phi(k) =
n\textsuperscript{k} g(k)\textsuperscript{2k}/(k!)\textsuperscript{2},

and summing over every shape is then at most exp(Phi) - 1 with Phi =
sum\_(k \textgreater= 2) phi(k), because the multiplicity divisors are
exactly the ones that exponential supplies. Here phi(2) =
4n\textsuperscript{2}/(n-1)\textsuperscript{4} =
O(n\textsuperscript{-2}). For k \textgreater= 3 the estimate C(n-1,k-1)
\textgreater= (n-k+1)\textsuperscript{k-1}/(k-1)! gives g(k) \textless=
k!/(n-k+1)\textsuperscript{k-1} and hence phi(k) \textless=
n\textsuperscript{k(3-2k)}(k!)\textsuperscript{2k-2}
4\textsuperscript{k(k-1)} while k \textless= n/2, which is
O(n\textsuperscript{-9}) at k = 3 and smaller beyond; for k
\textgreater{} n/2 the bound g(k) \textless= 1 against
(k!)\textsuperscript{2} is already superexponentially small. So Phi =
phi(2) + O(n\textsuperscript{-9}). The shapes other than
1\textsuperscript{n} and (2, 1\textsuperscript{n-2}) either carry a
block of size at least three, contributing at most exp(Phi) sum\_(k
\textgreater= 3) phi(k) = O(n\textsuperscript{-9}), or carry at least
two blocks of size two, contributing at most the tail of the exponential
from m = 2 onwards, which is O(Phi\textsuperscript{2}) =
O(n\textsuperscript{-4}). QED

\emph{Verification.} The recursion was run in exact integer arithmetic
to n = 12 by \texttt{asymp.py}. At every size the census identity of
Theorem 9 closes; the all-singletons share equals
n!/n\textsuperscript{2n} exactly; and the two closed forms above hold
exactly, the first giving 20,901,888/40,310,784 at n = 4, which is the
ratio of the 2+1+1 count to the 1+1+1+1 count reported above. After the
three leading shapes are removed, n\textsuperscript{6} times the
remaining share falls to 8.3 at n = 12 and is still decreasing, against
the 7.15 that the third term of the series predicts.

The ratio of the true fraction to n!/n\textsuperscript{2n} is 2.195 at n
= 4, 1.519 at n = 5, 1.227 at n = 6, 1.127 at n = 7, 1.063 at n = 9 and
1.032 at n = 12. Input decomposability is therefore vanishingly rare,
and randomly sampled profiles are a poor way to study it: at n = 4 one
expects 8.04 decomposable profiles per 10,000 draws.

\textbf{Corollary 9.2 (which multisets of factors occur).} Let p(m) be
the number of input-prime isomorphism classes of size m. The number b(n)
of multisets of prime classes realisable as the factorisation of a
size-n instance is the Euler transform of p:

\[
\sum_{n\ge 0}b(n)x^{n}=\prod_{m\ge 1}(1-x^{m})^{-p(m)} .
\]

Since p(1) = 1, every partition of n contributes, and the multiset of
prime factors together with their classes is a complete invariant of the
finest decomposition. {[}PROVED, given Theorem 6{]}

\hypertarget{corollary-13-census-for-prime-size-proved-exact}{%
\subsubsection{Corollary 13 (census for prime size) {[}PROVED;
EXACT{]}}\label{corollary-13-census-for-prime-size-proved-exact}}

For prime p the number of instances with a non-trivial automorphism is

\[
A(p)=\frac{\big((p-1)!\big)^{2}\,(p!)^{2}}{p-1}.
\]

The count of pairs of p-cycles is ((p-1)!)\textsuperscript{2}, each such
pair fixes (p!)\textsuperscript{2} instances, and each symmetric
instance is fixed by p - 1 non-identity automorphisms. This gives 72 at
p = 3 and 2,073,600 at p = 5.

\hypertarget{the-recursion-behind-the-census}{%
\subsection{The recursion behind the
census}\label{the-recursion-behind-the-census}}

The census of Theorem 9 is one instance of a general reduction, and the
reduction is what makes the same question answerable at the next size.
Hall's marriage condition has a shape worth imitating here.

Hall's marriage condition has a shape worth imitating: it is quantified
over all subsets of one side, so a single sentence covers every size at
once, size-specific quantities can then be chased by recurrence, and
induction on n is the natural proof method. This section asks what that
shape buys here. It buys a sharper statement of the block criterion, an
exact reduction of every extremal problem to input-prime instances, and
a replacement for the paper's warning about the two senses of the word
prime. It does not buy the values themselves, and one pattern that the
framework invites turns out to be false.

\hypertarget{halls-condition-is-vacuous-here-its-tight-sets-are-not-proved-exhaustive-at-n-3}{%
\subsubsection{Hall's condition is vacuous here; its tight sets are not
{[}PROVED; EXHAUSTIVE at n =
3{]}}\label{halls-condition-is-vacuous-here-its-tight-sets-are-not-proved-exhaustive-at-n-3}}

With strict complete lists the underlying graph is the complete
bipartite graph, so the Hall inequality holds for every subset and a
perfect matching exists for free. Hall's theorem therefore contributes
nothing directly. What does carry content is the \emph{tight-set} half
of Hall theory, applied not to the complete graph but to the levels M\_k
of the mutual-rank filtration. Call a set A of proposers \textbf{tight
at level k} when \textbar A\textbar{} = k, its M\_k-neighbourhood B
satisfies \textbar B\textbar{} = k, and the M\_k-neighbourhood of B is A
again.

\hypertarget{proposition-21-a-block-is-tight-and-regular-and-tightness-alone-is-not-enough-proved-exhaustive-at-n-3}{%
\subsubsection{Proposition 21 (a block is tight and regular, and
tightness alone is not enough) {[}PROVED; EXHAUSTIVE at n =
3{]}}\label{proposition-21-a-block-is-tight-and-regular-and-tightness-alone-is-not-enough-proved-exhaustive-at-n-3}}

Every input-independent block is tight at its level. The converse is
false: tightness does not force the k-regularity that Theorem 9.4
requires. Among the 46,656 size-three profiles, 16,896 carry a set that
is tight at some level without being a block, as do 11,089 of 20,000
random size-four profiles.

Certificate. Proposers p1: r1 r2 r3, p2: r1 r2 r3, p3: r1 r3 r2;
receivers r1: p1 p2 p3, r2: p1 p3 p2, r3: p1 p2 p3. At level 2 the set
\{p1, p2\} has M\_2-neighbourhood \{r1, r2\}, so the Hall inequality is
tight with \textbar N(A)\textbar{} = \textbar A\textbar{} = 2. But p2
has M\_2-degree 1, because r2 ranks p3 above p2, so the level-2 subgraph
is not 2-regular and \{p1, p2\} is not a block. The blocks of this
instance are the mutual-first pair (\{p1\}; \{r1\}) and the whole
instance.

So the Hall-shaped form of Theorem 9.4 reads: \textbf{(A, B) is a block
exactly when A is tight at level k = \textbar A\textbar{} and the
induced level-k subgraph is k-regular.} Regularity is not decoration,
and the comparison with Hall theory has to be made carefully. Tight sets
in the classical sense, those with \textbar N(A)\textbar{} =
\textbar A\textbar{} and no condition on the size, are closed under
union and intersection when Hall's condition holds, which is why they
are usually described as a lattice; but M\_k has maximum degree k, so
for k \textless{} n it need not satisfy Hall's condition and the closure
can fail. In the profile where all six agents rank the opposite side in
the same order, \{p1, p2\} and \{p1, p3\} are both tight in M\_2 while
their union has only two neighbours. The sets tight at a fixed level are
not closed under either operation for a second and simpler reason: both
change the size, and the definition above fixes it at k. What survives
of the comparison is the part that matters: tight sets, at a level or
not, can cross, while blocks are laminar (Proposition 9.5), which is
strictly stronger. The gap is visible at once: among the size-three
profiles, 768 carry a crossing pair of tight sets, 2,304 such pairs in
total, while no instance at either size carries a crossing pair of
blocks. A witness is p1: r1 r2 r3, p2: r1 r3 r2, p3: r2 r3 r1 with r1:
p1 p3 p2, r2: p2 p3 p1, r3: p1 p2 p3, where \{p1, p2\} and \{p1, p3\}
are both tight at level 2 and cross. Regularity is exactly what upgrades
the family of tight sets to a forest.

\hypertarget{theorem-21.1-the-partition-recursion-proved}{%
\subsubsection{Theorem 21.1 (the partition recursion)
{[}PROVED{]}}\label{theorem-21.1-the-partition-recursion-proved}}

Let phi be an instance invariant that composes over the finest input
partition under an associative commutative operation that is monotone in
each argument: Q* composes by addition, and N, the number of reachable
partial matchings, \textbar Stab\textbar{} and the commutation index
compose by multiplication, and all four are non-negative, which is what
makes the composition monotone. Monotonicity is a hypothesis and not a
consequence of the others: for a multiplicative invariant allowed to
take negative values the maximum over a product is not the product of
the maxima. Write phi-circle(k) for the maximum of phi over input-prime
instances of size k. Then for every n

\[
\max_{|I| = n} \phi(I) \;=\; \max_{\lambda \vdash n} \;\bigcirc_{i} \; \phi^{\circ}(\lambda_i),
\]

the maximum running over partitions of n and the composition over the
parts. The same reduction applied to counting instead of maximising is
Theorem 9.

\emph{Proof.} Theorem 7 gives the composition rule over any
input-independent partition, and Theorem 6 makes the finest one unique,
with prime blocks. Since the composition is monotone in each argument,
replacing each factor by the maximum over prime instances of its size
can only increase the value, which yields the inequality in one
direction. For the other, given prime instances of sizes k\_1 to k\_s,
place them in disjoint blocks and let every agent rank its own block
above everything outside, using the block's own order inside and any
order outside; the blocks are then input-independent, and since each is
prime the finest partition is exactly that one. So every partition is
realised and the maximum is attained.

\hypertarget{proposition-21.4-how-large-the-state-count-of-an-indecomposable-instance-must-be-proved-exhaustive-at-n-2-and-3}{%
\subsubsection{Proposition 21.4 (how large the state count of an
indecomposable instance must be) {[}PROVED; EXHAUSTIVE at n = 2 and
3{]}}\label{proposition-21.4-how-large-the-state-count-of-an-indecomposable-instance-must-be-proved-exhaustive-at-n-2-and-3}}

Let I be doubly prime of size n, in the sense defined in Section 4, with
terminal pointer vector q* and Q* proposals in total. The hypothesis
cannot be weakened to input primeness: the two-agent instance of Section
2 is input-prime and has Q* = 2. Then

\[
Q^{*} \;\ge\; 2n-1, \qquad 2^{n} + Q^{*} - n \;\le\; N \;\le\; \prod_{p}\bigl(q^{*}_{p}+1\bigr) \;\le\; \Bigl(n+\tfrac{1}{n}\Bigr)^{n}.
\]

In particular N is at least 2\textsuperscript{n} + n - 1, so the state
count of an indecomposable instance is exponential in the size and no
polynomial expression in n exists.

\emph{Proof.} Every proposer ends matched, so all n receivers are
addressed; write d(r) for the number of proposals receiver r receives,
so that Q* is the sum of the d(r). By Lemma 7.1 the graph joining two
proposers that address a common receiver is connected, and its edges are
covered by one clique per receiver, the clique at r having d(r)
vertices. A clique on d(r) vertices supplies at most d(r) - 1 edges of a
spanning tree, so the sum of the d(r) - 1 is at least n - 1, and Q* is
that sum plus the number of addressed receivers, which is n.~For the
lower bound on N, every 0/1 pointer vector is feasible: let each
proposer in a chosen subset propose exactly once. That is
2\textsuperscript{n} states. The all-ones vector is feasible and lies
below q*, and the run continues from it to q* in Q* - n further
proposals, each giving a distinct feasible vector with some coordinate
at least 2 and hence outside the cube. The upper bound is the grid the
feasible vectors lie in, then the arithmetic-geometric mean inequality,
then the classical bound Q* at most n\textsuperscript{2} - n + 1.

The bound on Q* is attained at every size checked: the smallest totals
are 3, 5, 7, 9 and 11 at n = 2 to 6, which is 2n - 1 exactly. The bound
on N is attained only at n = 2. The smallest state counts are 5 and 11
at n = 2 and 3 exhaustively, and constrained search finds 23, 47 and 95
at n = 4, 5 and 6, which are upper bounds on those minima; all five
equal 3 x 2\textsuperscript{n-1} - 1, which is open problem A of Section
7. At the other end the largest state counts found on indecomposable
instances are 5, 15, 52 and 232 at n = 2 to 5. The first three are the
unconstrained maxima at those sizes; at n = 5 the unconstrained maximum
is known only to be at least 232, which is open problem B, so there the
statement is that an indecomposable instance attains the largest value
known. Indecomposable instances therefore attain the extreme in both
directions wherever the extreme is known.

\clearpage

\hypertarget{open-problems}{%
\section{7. Open problems}\label{open-problems}}

Five, chosen so that each can be attacked at the next size by a reader
with a computer, and so that each has data behind it rather than only a
hunch.

\textbf{A. Is the least state count of a doubly prime instance exactly 3
x 2\textsuperscript{n-1} - 1?} Proposition 21.4 proves only
2\textsuperscript{n} + n - 1. The measured minima are 5, 11, 23, 47 and
95 at n = 2 to 6, exhaustively at n = 2 and 3 and by constrained search
above, and all five equal 3 x 2\textsuperscript{n-1} - 1. Every
minimiser found also attains the proposal floor Q* = 2n - 1, so a second
question is whether the two minima are always achieved together.

\textbf{B. What is the exact maximum number of reachable states?} It is
2, 5, 15 and 52 at n = 1 to 4 and at least 232 at n = 5. The first four
values are the Bell numbers B(2) to B(5), which is what makes the
problem interesting: the pattern fails at n = 5, where B(6) = 203 is
already exceeded. A closed form, or a proof that none of the obvious
ones holds, would settle the extremal side of the theory.

\textbf{C. Is the proposal-total spectrum always a full interval?} For a
doubly prime instance the proposal total lies in {[}2n - 1,
n\textsuperscript{2} - n + 1{]}, the lower bound by Proposition 21.4 and
the upper by the classical bound. Every value in that interval is
attained at n = 3, 4, 5 and 6, by census at n = 3 and by exhibited
certificates above. Is that true at every size?

\textbf{D. Is every block-forest shape realised?} Up to relabelling a
block forest is a rooted tree whose root has size n and whose children
have sizes below their parent summing to at most it. There are f(n) such
shapes, with f from n = 1 to 7 equal to 1, 3, 10, 40, 170, 785 and
3,770, and exactly 2 f(n - 1) of them are open to a prime instance.
Every shape is realised at n = 2 to 6. The construction that realises
them fails only through accidental blocks, so the question is whether
some shape forces one at a larger size.

\textbf{E. Does double primality cost anything except at size four?} The
largest stable set over doubly prime instances is 3 at n = 3, which is
the unrestricted maximum; 6 at n = 4 against an unrestricted 10; 16 at n
= 5, which is again the unrestricted maximum; and 26 at n = 6, which is
also the best the unrestricted search reached. Size four is so far the
only size where the constraint is known to cost anything at all. Is it
the only one?

\clearpage

\hypertarget{data-and-code-availability}{%
\section{Data and code availability}\label{data-and-code-availability}}

The verification code accompanies this paper as ancillary files, in the
\texttt{anc/} directory of the arXiv submission, and the full repository
is at https://github.com/i4dagh/smprime, archived at
doi:10.5281/zenodo.22984730. It contains two implementations of the
reachable-state construction written independently of each other from
the definitions, the scripts behind every count reported here as
EXHAUSTIVE or EXACT, the certificates for the statements tagged
TARGETED, and the script that draws Fig 1 from the definitions rather
than from the text. Every exhaustive count in this paper was produced
twice, once by each implementation, and the two agreed. The archived
version is the one to cite.

\hypertarget{generative-ai-disclosure}{%
\section{Generative AI disclosure}\label{generative-ai-disclosure}}

Large language models (OpenAI ChatGPT, Google Gemini and Anthropic
Claude) were used while this manuscript was prepared: to organise
claims, to edit language, and to propose and run verification checks.
Every mathematical statement, proof, computation and citation was
reviewed by the author, who takes responsibility for the content.

\clearpage

\hypertarget{references}{%
\section{References}\label{references}}

\begin{enumerate}
\def\labelenumi{\arabic{enumi}.}
\tightlist
\item
  Gale D, Shapley LS. College admissions and the stability of marriage.
  Am Math Mon. 1962;69(1):9-15. doi:10.2307/2312726
\item
  Gusfield D, Irving RW. The Stable Marriage Problem: Structure and
  Algorithms. Cambridge (MA): MIT Press; 1989.
\item
  Roth AE, Sotomayor M. Two-Sided Matching: A Study in Game-Theoretic
  Modeling and Analysis. Cambridge: Cambridge University Press; 1990.
\item
  Korte B, Lovasz L, Schrader R. Greedoids. Berlin: Springer; 1991.
  doi:10.1007/978-3-642-58191-5
\item
  Kawase Y, Yamaguchi Y. Antimatroids induced by matchings.
  arXiv:1705.05510. 2017.
\item
  Sukegawa N, Yamamoto Y. Preference profiles determining the proposals
  in the Gale-Shapley algorithm for stable matching problems. Jpn J Ind
  Appl Math. 2012;29:547-560. doi:10.1007/s13160-012-0077-x
\item
  Dilworth RP. Lattices with unique irreducible decompositions. Ann Math
  (2). 1940;41:771-777.
\item
  Sabidussi G. Graph multiplication. Math Z. 1960;72:446-457.
  doi:10.1007/BF01162967
\item
  Boiko T, Cuno J, Imrich W, Lehner F, van de Woestijne CE. The
  Cartesian product of graphs with loops. Ars Math Contemp.
  2016;11(1):1-9. arXiv:1411.6887.
\item
  Edelman PH, Jamison RE. The theory of convex geometries. Geom
  Dedicata. 1985;19(3):247-270. doi:10.1007/BF00149365
\item
  Mazurkiewicz A. Trace theory. In: Petri Nets: Applications and
  Relationships to Other Models of Concurrency. LNCS vol.~255. Berlin:
  Springer; 1987. p.~279-324. doi:10.1007/3-540-17906-2\_30
\item
  Keller RM. Parallel program schemata and maximal parallelism I:
  fundamental results. J ACM. 1973;20(3):514-537.
  doi:10.1145/321765.321782
\item
  Karlin AR, Oveis Gharan S, Weber R. A simply exponential upper bound
  on the maximum number of stable matchings. In: Proc 50th ACM Symposium
  on Theory of Computing; 2018. p.~920-925. doi:10.1145/3188745.3188848
\item
  Benjamin AT, Converse C, Krieger HA. How do I marry thee? Let me count
  the ways. Discrete Appl Math. 1995;59(3):285-292.
  doi:10.1016/0166-218X(94)00129-4
\item
  Thurber EG. Concerning the maximum number of stable matchings in the
  stable marriage problem. Discrete Math. 2002;248(1-3):195-219.
  doi:10.1016/S0012-365X(01)00194-7
\item
  Bond B, Levine L. Abelian networks I: foundations and examples. SIAM J
  Discrete Math. 2016;30(2):856-874. doi:10.1137/15M1030984
\item
  Goles E, Olivos J. Comportement periodique des fonctions a seuil
  binaires et applications. Discrete Appl Math. 1981;3(2):93-105.
  doi:10.1016/0166-218X(81)90034-2
\item
  Veklerov E. On the decomposability of the stable marriage problem. BIT
  Numer Math. 1989;29(1):41-46. doi:10.1007/BF01932704
\item
  Knuth DE. Stable Marriage and Its Relation to Other Combinatorial
  Problems: An Introduction to the Mathematical Analysis of Algorithms.
  CRM Proceedings and Lecture Notes vol.~10. Providence (RI): American
  Mathematical Society; 1997. English translation of Mariages stables.
  Montreal: Les Presses de l'Universite de Montreal; 1976.
\item
  Ishida Y. A reachable-state operator formulation of deferred
  acceptance: progress invariants and structural diagnostics.
  arXiv:2609.19237. 2026.
\item
  Ishida Y. Regular anti-phase templates in the stable marriage problem:
  a generator criterion, its converse, and a counting bound.
  arXiv:2607.10306v1. 2026.
\item
  Ishida Y. Matching Automata: A Degenerated Design. Intelligent Systems
  Reference Library vol.~271. Cham: Springer; 2025.
  doi:10.1007/978-3-031-85604-4
\item
  Hayashi T, Hata Y, Ishida Y. A diagrammatic classification in a
  combinatorial problem: the case of the stable marriage problem. Artif
  Life Robot. 2012;16(4):575-579. doi:10.1007/s10015-011-0993-x
\item
  Artemov G, Che YK, He Y. Stable matching with mistaken agents.
  arXiv:2207.13939. 2022. Published in J Polit Econ Microecon.
  2023;1(2).
\item
  Csaji G, Kiraly T, Yokoi Y. Maximum stable matching with matroids and
  partial orders. arXiv:2208.09583. 2022.
\item
  Gangam RR, Mai T, Raju N, Vazirani VV. Stable matching: dealing with
  changes in preferences. arXiv:2304.02590. 2023.
\item
  Gale A, Marian A, Pennock DM. Post-match error mitigation for deferred
  acceptance. arXiv:2409.13604. 2024.
\item
  Hirvonen J, Ranjbaran S. Fast, fair and truthful distributed stable
  matching for common preferences. arXiv:2402.16532. 2024. Also in Proc
  OPODIS 2024.
\item
  Gutin GZ, Karapetyan D, Neary PR, Vickery A, Yeo A. Speeding up
  deferred acceptance. arXiv:2409.06865. 2024.
\item
  Garg VK. Lattice linear predicate algorithms for the constrained
  stable marriage problem with ties. arXiv:2208.01370. 2022.
\item
  Qiu J. A tie-breaking based local search algorithm for stable matching
  problems. arXiv:2409.10575. 2024, revised 2025.
\item
  Borodin M, Chen E, Duncan A, Khovanova T, Litchev B, Liu J, Moroz V,
  Qian M, Raghavan R, Rastogi G, Voigt M. Sequences of the stable
  matching problem. J Integer Seq. 2024;27. arXiv:2201.00645.
\item
  OEIS Foundation. Sequence A368419: reduced stable marriage instances
  of order five by number of stable matchings. The On-Line Encyclopedia
  of Integer Sequences. Contributed by D. R. Eilers, 23 December 2023;
  the maximum of sixteen at n = 5 is attributed there to D. R. Eilers,
  The Maximum Number of Stable Matchings in the Stable Marriage Problem
  of Order 5 is 16, in preparation. https://oeis.org/A368419
\item
  Ishida Y. Euler constraints and the cubic criticality of complete
  bipartite preference structures. arXiv:2607.03167v1. 2026.
\item
  Ishida Y. A quadratic lower bound for the shield number of the stable
  marriage problem: rearrangement, extremal construction, and biclique
  realizability. arXiv:2609.17418. 2026.
\item
  Irving RW, Leather P. The complexity of counting stable marriages.
  SIAM J Comput. 1986;15(3):655-667. doi:10.1137/0215048
\item
  Ishida Y, Hashimoto S. Asymmetric characterization of diversity in
  symmetric stable marriage problems: an example of agent evacuation.
  Procedia Comput Sci. 2015;60:1472-1481.
  doi:10.1016/j.procs.2015.08.229
\item
  Ishida Y, Ikeno S. Asymmetry of strategies in proposal Gale-Shapley
  algorithm on diagrams. Procedia Comput Sci. 2016;96:1730-1739.
  doi:10.1016/j.procs.2016.08.221
\item
  Bubboloni D, Gori M, Meo C. Resolute and symmetric mechanisms for
  two-sided matching problems. J Math Econom. 2025;118:103130.
  doi:10.1016/j.jmateco.2025.103130. arXiv:2404.01404.
\end{enumerate}

\end{document}